\documentclass[twocolumn,trackchanges]{aastex631}

\newcommand{\ppl}{\left(}
\newcommand{\ppr}{\right)}

\newcommand{\brl}{\left[}
\newcommand{\brr}{\right]}

\newcommand\candsix{\text{D1005+68 }}
\newcommand\candseven{D1006+69 }
\newcommand\candeight{DWJ0954+6821 }
\newcommand\candnine{D1009+68 }
\usepackage{graphicx}
\usepackage{amsmath}
\usepackage{mathtools}
\usepackage{amssymb}
\usepackage{cancel}
\usepackage[normalem]{ulem}

\newcommand{\code}[1]{\texttt{#1}}

\usepackage{graphicx}

\shorttitle{Diverse Faint Dwarfs in M81}
\shortauthors{Gozman et al.}

\graphicspath{{./}{figures/}}
\begin{document}

\title{Exploring the Diversity of Faint Satellites in the M81 Group}

\correspondingauthor{Katya Gozman}
\email{kgozman@umich.edu}

\author[0000-0003-2294-4187]{Katya Gozman}
\affiliation{Department of Astronomy, University of Michigan, 1085 S. University Ave, Ann Arbor, MI 48109-1107, USA}

\author[0000-0002-5564-9873]{Eric F.\ Bell}
\affiliation{Department of Astronomy, University of Michigan, 1085 S. University Ave, Ann Arbor, MI 48109-1107, USA}

\author[0000-0002-2502-0070]{In Sung Jang}
\affiliation{Department of Astronomy and Astrophysics, University of Chicago, Chicago, IL 60637, USA}

\author[0009-0002-9085-5928]{Jose Marco Arias}
\affiliation{Department of Astronomy, University of Michigan, 1085 S. University Ave, Ann Arbor, MI 48109-1107, USA}

\author[0000-0001-6380-010X]{Jeremy Bailin}
\affiliation{Department of Physics and Astronomy, University of Alabama, Box 870324, Tuscaloosa, AL 35487-0324, USA}

\author[0000-0001-6982-4081]{Roelof S. de Jong}
\affiliation{Leibniz-Institut f\"{u}r Astrophysik Potsdam (AIP), An der Sternwarte 16, 14482 Potsdam, Germany}

\author[0000-0001-9269-8167]{Richard D'Souza}
\affiliation{Vatican Observatory, Specola Vaticana, V-00120, Vatican City State}

\author[0000-0001-9852-9954]{Oleg Y.\ Gnedin}
\affiliation{Department of Astronomy, University of Michigan, 1085 S. University Ave, Ann Arbor, MI 48109-1107, USA}

\author[0000-0003-2325-9616]{Antonela Monachesi}
\affiliation{Departamento de Astronom\'{i}a, Universidad de La Serena, Avda. R\'{a}ul Bitr\'{a}n 1305, La Serena, Chile}

%\author{David Nidever}
%\affiliation{Department of Physics, Montana State University, P.O. Box 173840, Bozeman, MT 59717-3840}
%\affiliation{National Optical Astronomy Observatory, %950 North Cherry Ave, Tucson, AZ 85719}

\author[0000-0003-0511-0228]{Paul A.\ Price}
\affiliation{Department of Astrophysical Sciences, Princeton University, Princeton, NJ 08544, USA}

\author[0000-0002-8406-0136]{Vaishnav V. Rao}
\affiliation{Department of Astronomy, University of Michigan, 1085 S. University Ave, Ann Arbor, MI 48109-1107, USA}

\author[0000-0003-2599-7524]{Adam Smercina}\thanks{Hubble Fellow}
\affiliation{Space Telescope Science Institute, 3700 San Martin Dr., Baltimore, MD 21218, USA}
\affiliation{Department of Astronomy, Box 351580, University of Washington, Seattle, WA 98195, USA}

%\author[0000-0002-0558-0521]{Colin Slater}
%\affiliation{Astronomy Department, University of Washington, Box 351580, U.W. Seattle, WA 98195-1580}

\begin{abstract}

In the last decade, we have been able to probe further down the galaxy luminosity function than ever before and expand into the regime of ultra-faint dwarfs (UFDs), which are some of the best probes we have of small-scale cosmology and galaxy formation. Digital sky surveys have enabled the discovery and study of these incredibly low-mass, highly dark-matter dominated systems around the Local Group, but it is critical that we expand the satellite census further out to understand if Milky Way and M31 satellites are representative of dwarf populations in the local Universe. Using data from HST/ACS, we present updated characterization of four satellite systems in the M81 group. These systems – D1005+68, D1006+69, DWJ0954+6821, and D1009+68 – were previously discovered using ground-based Subaru HSC data as overdensities in M81's halo and are now confirmed with HST/ACS by this work. These are all faint ($M_V \geq -7.9$) and consistent with old ($\sim$13 Gyr), metal-poor ([M/H] $< -1.5$) populations. Each system possesses relatively unusual features – including one of the most concentrated satellite galaxies with a Sérsic index of $n\sim5$, one of the most elliptical galaxies outside the Local Group with an $\epsilon \sim 0.6$, and one of the most compact galaxies for its magnitude. Two of the satellites have very low surface brightness, lower than most known galaxies in this absolute magnitude range. This work previews the scientific promise of the upcoming Rubin Observatory and Roman Telescope for illuminating the diversity of UFDs in the Local Volume and beyond.

\end{abstract}

\keywords{galaxies, ultra-faint dwarfs, M81 group}

\section{Introduction}\label{sec:intro}

%\begin{itemize}
    %\item UFDs are tiny, at the extreme of galaxy properties, useful probes of small-scale cosmology
    %\item gravitational wells to retain SN ejecta, fossil record of enrichment.
    %\item luminosity fxn
    %\item sky surveys accelerated UFD finding, resolved stars let us probe their stellar populations in greater detail
    %\item M81 group satellite history
%\end{itemize}

Hidden at the lowest end of the galaxy luminosity function are some of the best laboratories we have for studying galaxy formation and small-scale cosmology: ultra-faint dwarfs (UFDs). These systems are at the extremes of galaxy properties in many regards: they have very low stellar masses ($M_* \sim 10^5 M_\odot$) and are extremely faint, characterized with $M_{\rm V}>-7.7$ \citep{bullockboylankolchin2017, simonreview}, or $\mu_V = 26$ mag/arcsec$^2$ \citep{BELOKUROVEvans2022}. They also appear uniformly old and quite metal-poor, indicating that they have gone through very little chemical evolution. 

The low masses of these galaxies mean that they were susceptible to reionization, which could heat gas out of small dark matter (DM) halos and halt star formation \citep{brown2012}. Due to their shallow gravitational potentials, UFDs are extremely sensitive to strong supernova feedback, tidal and ram pressure stripping, and photoionization, all of which can drive outflows of gas and metals and quench star formation \citep{weiszboylankolchin2017}. This makes UFDs important for constraining galaxy formation physics in cosmological simulations %them powerful constraints on cosmological models 
\citep{bullockboylankolchin2017}. These constraints manifest themselves in two ways: 1) star-formation history (SFH) and metallicity diagnostics enable us to probe gas-driven physics, and 2) structural diversity informs us about gravitationally-driven physics in the low-mass regime.
% Bullock and Boylan-Kolchin 2017 + others.

The SFHs and metallicities of UFDs are extremely informative about reionization, environment, and feedback.
Very broadly, it appears that star formation in most UFDs was quenched 10 or more Gyr ago \citep[e.g.,][]{brown2012}. 
It is also clear that environment plays an important role. All UFDs close to their host are quenched \citep[e.g.,][]{slater2014ApJ_massdepdwarfquench, 
weisz2015ApJ_quenching, wetzel2015ApJ_envquenching, fillingham2016MNRAS_quenching}. While some isolated UFDs are quenched (e.g., Tucana B, Leo M and Leo K; \citealt{sand2022_tucanaB,mcquinn2023_leok_leom}), others continue to form stars or have had extended star formation histories (e.g., Leo T and Pegasus W; \citealt{irwin2007_leoT,mcquinn2023}). It therefore appears that star formation is possible in very low mass and isolated galaxies, but not inevitable, giving valuable insight into the complex interplay of stellar feedback, reionization and environment. Models have differing interpretations of this complex behavior. For example, UFDs in the GIZMO/FIRE suite of extremely high-resolution simulations of \cite{wheeler2019MNRAS_quenched_sims_FIRE} are all uniformly quenched early at $z \gtrsim 2$. On the other hand, \cite{rey2020MNRAS_edgeSF} are able to reproduce galaxies like Leo T in the EDGE simulations, creating low-mass field dwarfs that reignite and sustain late-time SF due to competition between stellar feedback and mass growth, despite being sufficiently low mass to be quenched by reionization.

\cite{munshi2019ApJ_SFcosmo} shows that varying SF models in cosmological simulations changes the number, stellar mass, and metallicities of satellites around dwarf galaxies by substantial factors. \cite{agertz2020} explores the role that subgrid physics has on various dwarf scaling relations by simulating a single halo mass $10^9 M_\odot$ dwarf and varying the feedback and resolution. While most scaling relations do not depend on the model used, they find that the mass--metallicity relation is the only one that can discriminate between different subgrid models, as the simulated galaxies vary on 4 orders of magnitude in metallicity depending on their levels of feedback. 

As we have probed deeper toward the faintest end of the satellite luminosity function, we have discovered a zoo of dwarf and ultra-faint dwarf galaxies. Not only are they unique in their structural and chemical properties, but their structural diversity is also a crucial tracer of gravitationally driven physics. Being at the lowest end of the luminosity function make UFDs excellent probes of dark matter halo structure. Recently, \cite{errani2024ApJ_UMIIIUniI_DMdom} report that the newly discovered Ursa Major III/UNIONS 1 (UMa3/U1) system is likely the most dark-matter dominated galaxy ever found, requiring it to reside in a dense DM cusp in order to not be stripped by the MW tidal field. Studies of UFDs have also helped constrain the nature of dark matter itself. 
\cite{esteban2023_darkmatterpower} and
\cite{kim2021arXiv_sat_velfunc_DM} use velocity information from galaxies including UFDs to constrain the power spectrum and properties of dark matter, while \cite{nadler2020ApJ_galaxyhaloconnection} show how UFDs constrain small-scale structure abundances and DM microphysics and \cite{koulen2024_eriII_dmblackholes} use Eridanus II and its nuclear star cluster to understand the nature of dark matter candidates.

While galaxies such as UMa3/U1 are extremely compact, other UFDs are faint but quite extended.  Crater II has an extremely low velocity dispersion and defies expectation as a very large but very diffuse dwarf, one of the lowest surface brightness galaxies ever found \citep{torrealba2016_crater2}. Simulations show that by varying the accretion history and timing of dynamical mass growth of field UFDs, one can produce a 1 dex scatter in present-day stellar mass at fixed halo mass. Galaxies with more delayed mass build-up are fainter and more diffuse, and extremely low surface brightness and highly diffuse galaxies can be created through late dry mergers and accretion of ex-situ stars \citep{rey2019ApJ_scatterUFDmassSB}. 

This also points to dwarf mergers as a potential nonnegligible driver of satellite structure. Simulations are able to produce UFDs with spatially extended stellar populations resembling a stellar halo \citep{tarumi2021_tucanaII}, which has been seen observationally in a number of UFDs \citep{chiti2021NatAs_tucII, tau2024_UFD_stellarhalos}. One of the most well known is Tucana II \citep{chiti2021NatAs_tucII}, an extremely metal-poor system with an extended stellar halo that arose from either early merger or strong bursty feedback. Combining predictions from hydrodynamical and DM-only simulations, \cite{revaz2023A&A_compactUFD} find that initially disconnected stellar building blocks at high redshift can merge and create systems that are not only naturally extended with signatures of a stellar halo, but also moderately elliptical, which could erroneously be interpreted as a indication of tidal interaction. A scenario like this has been hypothesized for Bo\"otes I, a MW satellite with extremely high ellipticity ($\epsilon = 0.68 \pm 0.13$) and two distinct stellar populations that could have arisen from the merger of two building block systems \citep{frebel2016ApJ_bootesIbuildingblocks, longeard2022_bootesI}.

In fact, a number of MW satellites with pronounced elongations have been argued to either be or have been under tidal influence. The unusual kinematics and large size of low-density and low surface brightness satellites such as Crater II and Antlia II have led to theories that they have been tidally stripped by the MW \citep{sanders2018_crater2,fu2019_crater2hercules, torrealbe2019_antlia2}. Tucana III \citep{li2018_tucIII} is a confirmed tidally disrupted dwarf galaxy, showing distinct tidal tails. Cetus III \citep{homma2018_cetusIII}, Bo\"otes I \citep{longeard2022_bootesI}, and Hercules \citep{coleman2007_hercules, munoz2018_hercules} are some of the most elongated satellites ever found, with ellipticities $\geq0.7$. The tidal influence on Hercules specifically is contentious. Though it was found to be embedded in a stellar stream \citep{sand2009ApJ_hercules}, an updated study of stars in its outskirts found that it does not have a significant velocity gradient as previously thought \citep{aden2009ApJ...hercules_velgradient, longeard2023MNRAS_hercules}, making it a less likely candidate for tidal stripping.

%Some UFDs have properties that don't have analogs in other systems, such as Eridanus II \citep{bechtol2015_eriII,koposov2015_eriII,crnojevic2016_eriII,simon2021_eriII}, another isolated UFD that appears to be the lowest-mass galaxy hosting a central star cluster. Other galaxies are chemically peculiar, such as Reticulum II \citep{roederer2016_retII, ji2016_retII} and Tucana III \citep{hansen2017_tucIII}, which are the first satellites of their kind to show stars enriched in the r-process. 

A major roadblock is that while we observe these diverse structures, our current simulations are not reproducing them. \cite{richstein2024} show that galaxies generated by simulations of varied dark matter prescriptions, subgrid models, and feedback can reproduce the variations in half-light radius and luminosity that UFDs exhibit, but each individual simulation cannot reproduce the entire range of observed UFDs. Some simulation suites, notably FIRE-2 and TNG, produce galaxies much too diffuse, and in the case of FIRE-2, much fainter than we have observed. The wide variations in observed stellar structures of UFDs are crucial to constraining and illuminating the physical processes that govern our faintest neighbors.

Though UFDs were predicted to exist by the $\Lambda$CDM cosmological framework decades ago, we have only recently been able to detect and observe them with enough resolution to make substantial progress in sensitivity to model physics.
Wide-field imaging efforts such as the Sloan Digital Sky Survey (SDSS), Dark Energy Survey (DES), and the Pan-Andromeda Archaeological Survey (PAndAS) have enabled successful searches for our faintest neighbors and led to an exponential increase in the number of UFDs discovered over the last two decades. The bulk have been identified in our Local Group, with the MW census complete to M$_{\rm v}>-4$ and the M31 census complete out to M$_{\rm v}\sim -7$ \citep{mcconnahie2018, drlicawagner2020}. 

While the properties of our closest neighbors are known, there is a risk of overtuning simulations and models to match observations of just two galaxies in our Local Volume. Evidence points to these galaxies not being representative of the satellite population of MW-mass systems in both number and star-formation history. The satellite luminosity function for nearby galaxies varies by an order of magnitude  \citep{geha2017ApJ_sagaI, smercina2018_m94lonely,carlsten2021ApJ_lumFuncsDwarfs}. This scatter may be influenced by merger history, as it has been established that satellite delivery is an important consequence of merger (e.g. the MC's bringing satellites to the MW, \citealt{deason2015MNRAS_MCs, kallivayalil2018ApJ_lmcInfall, patel2020ApJ_MCs}, and many M31 satellites that were brought in through its dominant merger, \citealt{dsouza2021MNRAS_m31}). As UFD abundance is predicted to be correlated with host halo assembly history \citep{bose2020MNRAS_ufdAssemblyHistory} (which is a driver of satellite number), it is imperative that we are able to complete satellite inventories down to UFD luminosities to understand the scatter in the satellite luminosity function. Additionally, among classical dwarfs, ones belonging to the Milky Way and M31 are generally quiescent, while the ELVES and SAGA surveys of satellites around MW-mass analogs have found quenched fractions 1$\sigma$ lower than that of the Local Group \citep{mao2024arXiv_sagacensus}. 

This is in stark contrast to the M81 group. At a distance of $\sim$ 3.6 Mpc \citep{radburnsmith2011_m81dist}, it comprises three main galaxies --- M81, M82, and NGC 3077 --- that are strongly interacting (unlike the MW or M31). Not only is M81 in the beginning stages of merger, \citep{smercina2020_m81} unlike the LG, many of its satellites show signs of recent star formation \citep{chiboucas2009_094469, weisz2011ApJ_acsSFdwarf, okamoto2019_d100669}. This makes the M81 group a particularly tantalizing target of study.  The group is also home to over 30 known dwarf satellites, including tidal dwarfs. \cite{chiboucas2013} found and confirmed 14 of them using the Hubble Space Telescope (HST), and two more were found by \cite{okamoto2015_m81} and \cite{okamoto2019_d100669} using Subaru's Hyper Suprime Cam (HSC). An additional faint satellite, \candsix, was found by \cite{smercina2017_d100568}. Most recently, \cite{bell2022_0954} conducted a search through seven $\sim$1.5 deg$^2$ fields of archival HSC data to identify six dwarf satellite candidates, finding many of them clustering around NGC 3077, indicating a possible satellite-of-satellites scenario during past group infall.

In this paper, we confirm the faint galactic nature of four satellites in the M81 group using HST/ACS follow-up and derive their structural properties. %We confirm the discovery of a new UFD in the group, \candeight, and present a re-characterization of three previously-known satellites, now with space-based follow-up. 
We find that these four galaxies are diverse in their properties, and that they are among the faintest satellites ever discovered in the M81 group. The techniques used in this paper are important precursors to analyses that will be done with the Rubin and Roman telescopes, whose facilities will offer unparalleled discovery potential for ultra-faint satellite populations of MW-mass galaxies out to 10 Mpc in the Local Volume.

\section{Observations}\label{sec:obs}

\begin{figure*}
    \centering
    \includegraphics[width=\textwidth]{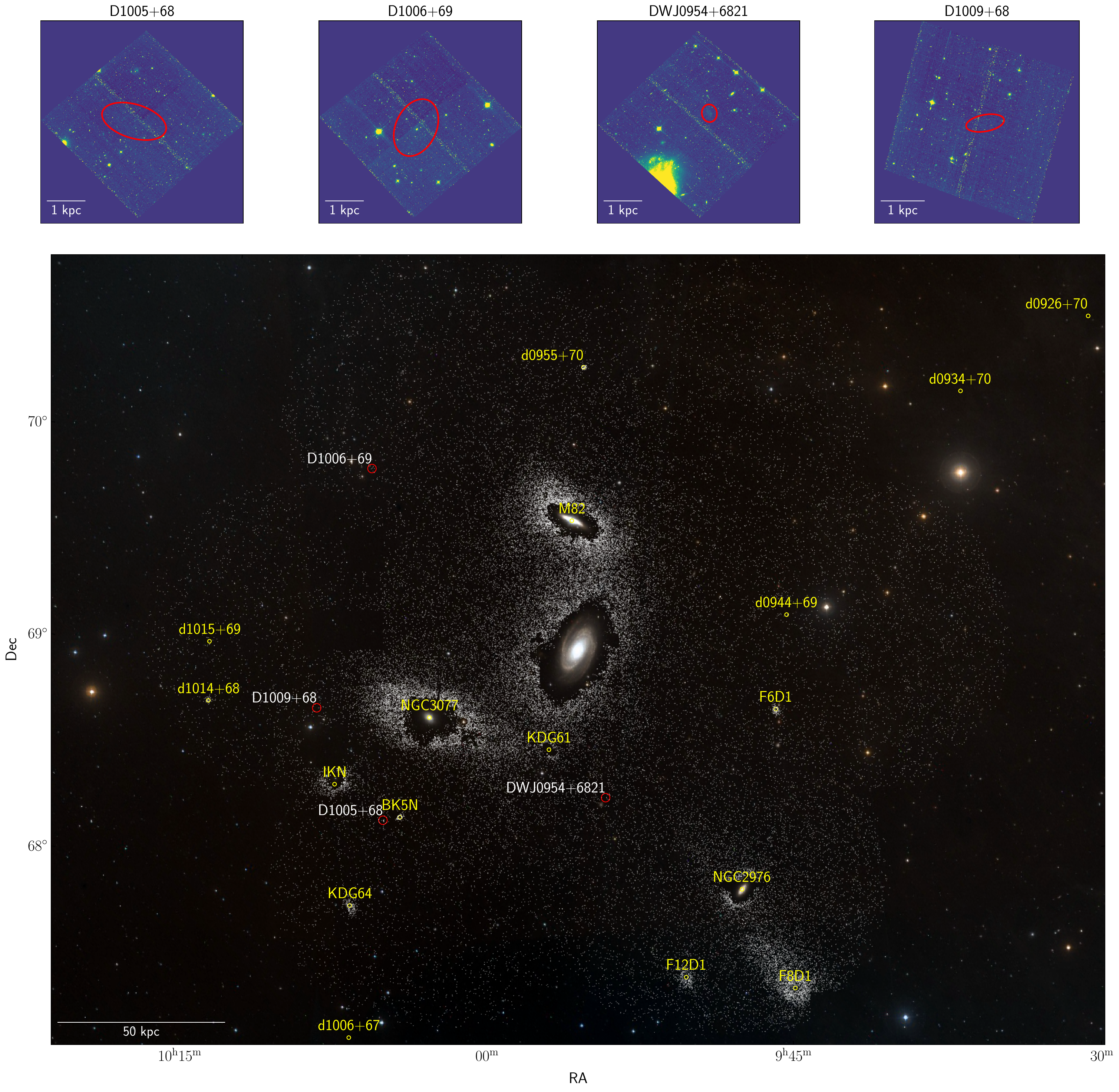}
    \caption{\textit{Top:} HST/ACS F814W images of our four dwarfs. The best-fit ellipses representing 2$a_h$ are shown in red for each galaxy. %\tbd{There's nothing obvious in these images. Can you play with the stretch to show the galaxies a bit better? Did you use a logarithmic/asinh stretch? - PAP.} \textbf{I'll be honest these galaxies don't really have anything in regards to diffuse brightness, so no matter what scalings I do you still can't really see anything in the images except for cand 8 (J0954) -- this one you can since a bit diffuse SB in it. My goal here is to just show the relative size and orientation of each candidate in relation to the M81 group to give a sense of scale, I don't expect anyone to see features -- KG}
    \textit{Bottom:} A map of the M81 group and select satellite galaxies. The background image is composed of images generated by Aladin \citep{aladin2000}, stacked to create an RGB frame using astropy's \texttt{make\_lupton\_rgb}. White dots represent stellar halo stars identified in Subaru/HSC data. Satellites highlighted in this work are shown as red dots with white text, other satellites are marked with yellow dots and yellow text.}
    %{\color{red} In Sung : Just a suggestion. I thought the upper 4 panels could be larger as in your poster, so readers can see the clustering of stars in the images. Alternatively, we could provide a background image in Fig. 6. You can use "open" circles to mark individual stars. This can be done after receiving the referee's report, if you prefer. {\color{blue} Katya: Thank you, I will do that after the referee's report in case they have other comments about the figure that need adjusting}} 
    \label{fig:m81_map}
\end{figure*}

The galaxies surveyed in this work have either been previously found in ground-based data or were identified as new candidates in \cite{bell2022_0954}. A map of the M81 system and a subset of its satellites are shown in Figure \ref{fig:m81_map}. All source images were obtained using HST's Wide Field Camera (WFC) channel of the Advanced Camera for Surveys (ACS) instrument \citep{ford1998_ACS} between Oct 2022-June 2023. Observations were taken in the F606W and F814W filters as part of HST SNAP 17158 (PI: Bell). All HST data used in this paper can be found in MAST: \dataset[10.17909/r3yg-dx15]{http://dx.doi.org/10.17909/r3yg-dx15}. The field of view is 202x202 arcsec, with plate scale of 0.05 arcsec/px. For D1005$+$68 and D1006$+$69, the satellites extended to large enough radius to benefit from our use of WFC3 parallels in F606W and F814W. Of note is that buffer constraints prevented our use of a cosmic ray split in F814W in the parallels; we identify stars in F606W only, and then use (the cosmic ray contaminated) F814W for color information. Full details of the observations can be found in Table \ref{tab:obs}.  

We reduced the data and performed PSF-fitting photometry using the software package \texttt{DOLPHOT} \citep{dolphot2000, dolphot2016} using TinyTim PSFs \citep{krist2011_tinytim}. Manual aperture corrections were applied to increase photometric accuracy, following the method described in \cite{jang2023_acsphotometry}.
Sources in ACS/WFC images were selected to be stars if they have 
%  & (asts['err']<=2) & (asts['err_1']<=2)  & (asts['type']==1)

\begin{enumerate}
    \item $\rm S/N>5$ in each passband
    \item $-0.06 < {\rm SHARPNESS}_{F606W} + {\rm SHARPNESS}_{F814W} < 1.3$
    \item ${\rm CROWDING}_{F606W}+{\rm CROWDING}_{F814W}<0.16$
\end{enumerate}

Sources in the parallel WFC3/UVIS2 fields were selected to be stars if they have

\begin{enumerate}
    \item $\rm S/N_{F606W} > 5.1$ and $\rm S/N_{F814W} > 3.2$
    \item $-0.19 < {\rm SHARPNESS}_{F606W} + {\rm SHARPNESS}_{F814W} < 1.5$
    \item ${\rm CROWDING}_{F606W}+{\rm CROWDING}_{F814W}<0.20$
\end{enumerate}

For both ACS and WFC3, we also only include stars with $\rm GF \neq 0$ (the star is not masked out by the extended source mask) with an error flag $\leq 2$ (few saturated pixels) and type $=1$ (clean point source). The bright star masks and parallel fields used are shown in Appendix \ref{masks}.

These cuts are the same as ones made in studies of faint RGB stars in crowded fields in works such as \cite{radburnsmith2011_m81dist} and \cite{jang2020_m101}.
%$S/N>4$ in each passband, reduced $\chi^2$ (the square of the deviation from a PSF, divided by the per-pixel uncertainties squared) less than 6.25, $|{\rm SHARPNESS}_{F814W}|<0.2$ and ${\rm SHARPNESS}_{F606W}^2 + {\rm SHARPNESS}_{F814W}^2 < 0.075$, ${\rm CROWDING}_{F814W}<0.1$ and ${\rm CROWDING}_{F606W}+{\rm CROWDING}_{F814W}<1.0$, {\bf Katya, why do we have a constraint on error in 606 and 814 being $<2$? I am confused} and ${\rm FLAG}=1$. {\bf Katya; why are these cut so much more ornate and complex than e.g., Jang 2021?}

%\textit{\textbf{Not sure, I'm not the one that made these cuts or constraints?? I just use sources that have FLAG = 1, which I thought was a flag In Sung made that incorporated all the cuts.--KG}}

We also ran a suite of artificial star tests (ASTs) in order to quantify the completeness of our sample. Between $\sim$ 476,000-477,600 artificial stars were drawn from a uniform grid of sources with magnitudes between $22<\text{F814W}<28$ and colors between $-0.5<\text{F606W-F814W}<2.5$, injected across the field of view.  

\begin{deluxetable*}{c c c c}

\tablecaption{Observation details for each dwarf satellite\label{tab:obs}}

\tablehead{
\colhead{Galaxy} & \colhead{Filter (Exp Time)} & \colhead{Num. Exposures} &  \colhead{50\%(80\%) Completeness}
}

\startdata 
D1005+68 & F606W ($460s+340s$) & 2 & 26.5(25.6)\\
 & F814W ($350s+340s$)  & 2 &  25.5(24.6) \\
D1005+68 Parallel & F606W ($190s+970s$) & 2 & --\\
 & F814W ($685s$)  & 1 &  -- \\
D1006+69 & F606W ($460s+340s$) & 2 & 26.6(25.6)\\
 & F814W ($350s+340s$) & 2 & 25.6(24.6) \\
D1006+69 Parallel & F606W ($190s+970s$) & 2 & --\\
 & F814W ($685s$)  & 1 &  -- \\
J0954+6821 & F606W ($470s+340s$) & 2 & 26.6(25.6) \\
 & F814W ($350s+340s$) & 2 & 25.6(24.6)\\
D1009+68 & F606W ($460s+340s$) &  2 & 26.5(25.5) \\
 & F814W ($350s+340s$) & 2 & 25.5(24.5) \\
\enddata

\end{deluxetable*}

The 50\% (80\%) completeness limits for each galaxy in each filter are cited in Table \ref{tab:obs}.  See Section \ref{sec:completeness} for further discussion of completeness in our samples. All source magnitudes are de-reddened according to the Galactic extinction maps from \cite{schlegel1998} and calibrated to the HST/ACS (or HST/WFC3 for parallel fields) filters using Table 6 of \cite{schlaflyfinkbeiner2011}, assuming a \cite{fitzpatrick1999} reddening law with $R_V = 3.1$. Using this, $A_{F606W}/E(B - V)_{SFD} = 2.471 (2.488)$ and $A_{F814W}/E(B - V)_{SFD} = 1.526 (1.536)$ for ACS(WFC3).

\section{Satellite Discovery}\label{sec:sat_disc}
\subsection{D1005+68}
D1005+68 was first discovered as a stellar overdensity by \cite{smercina2017_d100568} using the Subaru Hyper Suprime-Cam. They measured an absolute V-band magnitude of $M_V = -7.94^{+0.38}_{-0.50}$, making it one of the faintest confirmed UFDs outside the Local Group, and find that it is within the virial radius of M81 satellite BK5N, possibly making it a satellite-of-a-satellite. They measured a TRGB of $I_{TRGB} = 23.96^{+0.20}_{-0.25}$ using the technique in Appendix C of \cite{Monachesi2016} and fit PARSEC isochrones to get a best-fit metallicity of [Fe/H] $\sim -1.90$. This system was later followed up by \cite{okamoto2019_d100669}, with a good agreement with \cite{smercina2017_d100568}'s $M_V$ value. They found a half-light radius of $340 \pm 50$ pc, which differs significantly from \cite{smercina2017_d100568}'s value of $188^{+39}_{-41}$ pc, which they attribute to \cite{smercina2017_d100568}'s assumption of a circular stellar distribution.

\subsection{D1006+69}
D1006+69 was initially found by \cite{okamoto2019_d100669} during a visual examination for new stellar overdensities in a deep, photometric survey of resolved M81 stars using the Subaru Hyper Suprime-Cam. They derived its distance using the TRGB method, finding a transition in the luminosity function by applying a Sobel filter. They detected a transition at $i_0 = 24.64 \pm 0.08$ and conclude that it is one of the faintest and metal-poor dwarfs around M81, with an $M_V = -8.91 \pm 0.40$ and an estimated $[$M/H$]$ = $-1.83 \pm 0.28$.

\subsection{DWJ0954+6821}
DWJ0954+6821 was also discovered in Subaru/HSC imaging from \cite{okamoto2015_m81}'s dataset by \cite{bell2022_0954} by searching for overdensities in resolved stars. They found it has a high surface brightness (with detectable diffuse brightness) and is well-measured, with an $M_V = -7.1 \pm 0.25$. That said, it has partial crowding, leading to a sparser CMD. They find it is relatively compact, much like Pegasus V/Andromeda XXXIV or Tucana B. 

\subsection{D1009+68}
D1009+68 was also initially found in \cite{okamoto2019_d100669} using the same detection and distance methods as for D1006+69. They detected a Sobel filter transition at $i_0 = 24.47 \pm 0.12$ and found an $M_V = -8.73 \pm 0.45$ and estimated an $[$M/H$]$ = $-1.43 \pm 0.28$.

\begin{deluxetable*}{ll|llll}
%\begin{deluxetable*}{lllllllll}
\tablecaption{Observed and measured properties of dwarf satellite galaxies in M81\label{Table}}
\tablewidth{0pt}
\tablehead{
\colhead{Parameter} & \colhead{Unit} & \colhead{D1005+68$^a$} & \colhead{D1006+69$^b$} & \colhead{J0954+6821$^c$} & \colhead{D1009+68$^b$} }

\setlength{\tabcolsep}{10pt}
\renewcommand{\arraystretch}{1.5}

\startdata
RA (J2000) & - & 10$^h$05$^m$34.13$^{+1.07^s}_{-1.00}$ & 10$^h$06$^m$55.44$^{+0.25^s}_{-0.27}$ & $09^h54^m06.95_{-0.17}^{+0.19^s}$
& $10^h09^m14.10_{-0.48}^{+0.46^s}$ \\
Dec (J2000) & - &
$68^\circ14^m22.05_{-3.25}^{+2.78^s}$& $69^\circ54^m15.81_{-1.46}^{+1.34^s}$ & $68^\circ21^m50.86_{-1.03}^{+1.11^s}$ & $68^\circ45^m25.98_{-1.36}^{+1.43^s}$\\
$a_h$ & arcmin & 
$0.41^{+0.1}_{-0.1}$ & 
$0.37^{+0.3}_{-0.2}$ & 
$0.11^{+0.02}_{-0.02}$ &
$0.21_{-0.04}^{+0.06}$\\
$a_h$ (physical) & pc & 
$429_{-99}^{+137}$ &
$384_{-154}^{+359}$ &
$112_{-17}^{+21}$ &
$220_{-45}^{+60}$\\
$r_h$ & arcmin &
$0.29_{-0.05}^{+0.07}$ & 
$0.29_{-0.1}^{+0.3}$ & 
$0.10_{-0.01}^{+0.02}$ &
$0.14_{-0.02}^{+0.03}$\\
$r_h$ (physical) & pc & 
$307_{-57}^{+75}$ & 
$309_{-117}^{+272}$ & 
$102_{-14}^{+16}$ & 
$145_{-24}^{+31}$
\\
$\epsilon$ & - & 
$0.49_{-0.19}^{+0.13}$ &
$<0.53$ &
$<0.30$ & 
$0.57_{-0.19}^{+0.13}$ \\
P.A. & deg & 
$73_{-13}^{+13}$ &
$-25_{-22}^{+31}$ &
$7_{-50}^{+40}$ & 
$100_{-8}^{+9}$\\
S\'ersic index $n$ & - & 
- & $4.8_{-1.9}^{+2.1}$ & - & - \\
$\Delta$BIC$^d$ & - & 3.4 & -5.0 & 4.1 & 2.3 \\
$[$M/H$]$ & - & 
$-1.7_{-0.4 (\text{r}) - 0.3(\text{d}) -0.2(\text{a})}^{+0.3 (\text{r}) + 0.3(\text{d}) +0.2(\text{a})}$ &
$-1.5_{-0.1 (\text{r}) - 0.2(\text{d}) -0.2(\text{a})}^{+0.1 (\text{r}) + 0.2(\text{d}) +0.2(\text{a})}$ & 
$-1.7_{-0.3 (\text{r}) - 0.4(\text{d}) -0.2(\text{a})}^{+0.2 (\text{r}) + 0.2(\text{d}) +0.2(\text{a})}$ & 
$-2.2_{-0.04 (\text{r}) - 0.3(\text{d}) -0.2(\text{a})}^{+0.5 (\text{r}) + 0.3(\text{d}) +0.2(\text{a})}$  \\
$\rm M_V$ & mag & 
$-7.7_{-0.2 (\text{r}) - 0.2(\text{d})}^{+0.2 (\text{r}) + 0.3(\text{d})}$ & 
$-7.8_{-0.2 (\text{r}) - 0.3(\text{d})}^{+0.2 (\text{r}) +0.3(\text{d})}$ & 
$-7.9_{-0.1 (\text{r}) -0.2(\text{d})}^{+0.2 (\text{r})+0.3(\text{d})}$ & 
$-7.7_{-0.1 (\text{r}) -0.2(\text{d})}^{+0.2 (\text{r})+0.2(\text{d})}$ \\
$M_*$ & $M_\odot$ & 
$2.3_{-0.4(\text{r})-0.5(\text{d})}^{+0.3(\text{r})+0.5(\text{d})} \times 10^5$ &
$2.5_{-0.4(\text{r})-0.5(\text{d})}^{+0.5(\text{r})+0.7(\text{d})} \times 10^5$ &
$2.8_{-0.4(\text{r})-0.7(\text{d})}^{+0.3(\text{r})+0.7(\text{d})} \times 10^5$ &
$2.2_{-0.3(\text{r})-0.4(\text{d})}^{+0.2(\text{r})+0.5(\text{d})} \times 10^5$\\
$N_*$ & - &  
$22_{-3}^{+3}$ &
$24_{-4}^{+5}$ &
$30_{-4}^{+4}$ &
$20_{-2}^{+2}$\\
$R_{\text{tidal}}$ & pc & 
$\sim 1200$ & $\sim 1300$ & $\sim 600$ & $\sim 1000$\\ 
\enddata

% D1005+68 & 1 & $148.5292\pm0.0008$ & $68.3641 \pm 0.0003$ & $-7.94^{+0.38}_{-0.50}$ & 
% $188^{+39}_{-41}$ & b/a & pa & Fe/h & $5.40^{+0.22}_{-0.16}$ \\
% D1006+69 & 2 & $148.5292\pm0.0008$ & $68.3641 \pm 0.0003$ & $-7.1\pm0.25$ & 
% $78\pm8$ & $0.55 \pm 0.1$ & $25\pm 8$ & Fe/h  \\
% J0954+6821 & 3 & $148.5292\pm0.0008$ & $68.3641 \pm 0.0003$ & $-7.1\pm0.25$ & 
% $78\pm8$ & $0.55 \pm 0.1$ & $25\pm 8$ & Fe/h  \\
% d0944+69 & 4 & $148.5292\pm0.0008$ & $68.3641 \pm 0.0003$ & $-7.1\pm0.25$ & 
% $78\pm8$ & $0.55 \pm 0.1$ & $25\pm 8$ & Fe/h  \\

\tablecomments{First discovered in: $^a$\cite{smercina2017_d100568}, $^b$\cite{okamoto2019_d100669}, $^c$\cite{bell2022_0954}. $^d\Delta$BIC = BIC$_{\text{S\'ersic}}$ - BIC$_{\text{Exp}}$ A positive number means an exponential fit is favored over a S\'ersic fit. All physical properties are calculated assuming the distance of M81 = 3.6 Mpc; no uncertainty on distance has been propagated through. For values with multiple errors, we explicitly state the uncertainty from random error as (r), from distance error as (d), and from age error as (a). Distance uncertainties in $M_V$ and mass were calculated by redoing the magnitude and mass measurements using a mock stellar population shifted 400 kpc closer and further from the nominal distance of 3600 kpc. One can assume a 30\% uncertainty on age for the values of $M_V$ and mass as reported in \cite{harmsen2017}. The age uncertainty on $[$M/H$]$ was derived from calculating the difference in metallicity between an 8 and 13 Gyr isochrone. For D1009+68, because our best-fit metallicity runs into the lower bound of our interpolation grid, we take its upper uncertainty in distance and mirror it as the lower uncertainty as well.
}
\label{tab:mcmc}
\end{deluxetable*}

\section{Completeness}\label{sec:completeness}
\begin{figure}
    \centering
\includegraphics[width=0.48\textwidth]{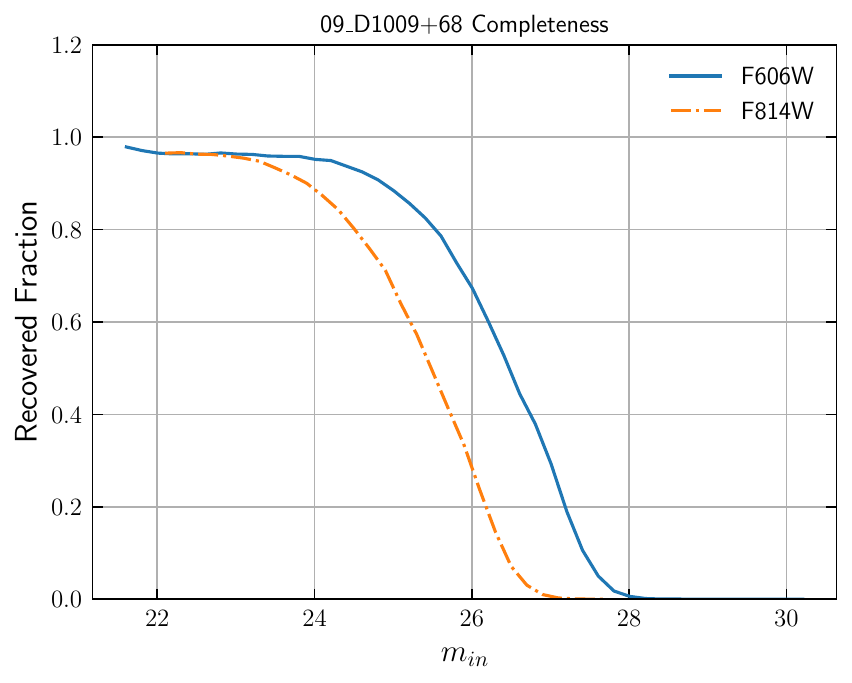}
    \caption{Completeness curve for one galaxy in our sample, D1009+68.}
    \label{fig:completness_curve}
\end{figure}

We compute two metrics of completeness for each galaxy using our ASTs: the overall 50\% and 80\% completeness levels, and also an individual completeness for each star. The former is calculated by binning the data in magnitude and calculating the ratio of recovered stars to total number of stars in each bin. A representative completeness curve is shown in Figure \ref{fig:completness_curve}. We then interpolate this curve to find the 50\% and 80\% completeness levels, which are cited for each galaxy in Table \ref{tab:obs}. To measure the completeness of each individual star, we create a 4D tuple that contains the color, magnitude, and X, Y positions of each star in our data and the ASTs. We loop through each star, find the 100 closest ASTs using the Mahalanobis distance metric, and calculate the percent of those 100 closest points that are recovered. This method has two advantages: the recovered fraction is calculated taking into account both color and spatial information, and the Mahalanobis distance prevents problems that the simple distance formula has with the large dynamic range of the 4D tuples (large X, Y coordinates compared to the small values of color). We find the 50\% and 80\% completeness limits for each filter to be almost identical among all images.

\section{Structural Parameters}\label{sec:struc_params}
\subsection{MCMC Fitting}
For each system, we estimated structural parameters such as the centroid $(x_0, y_0)$\footnote{The centroid is measured as an arbitrary offset from the center of each galaxy as cited previously in the literature}, ellipticity $\epsilon = 1 - \frac{b}{a}$, and half-light radius $r_h$ of each UFD in order to compare them to the broader population of galactic and extragalactic UFDs. We used a Markov Chain Monte Carlo (MCMC)-based maximum-likelihood approach following the work of \cite{martin2016} and \cite{mcquinn2023} that accounts for spatial incompleteness in the dataset. We assume that each galaxy contains $N_*$ stars whose surface brightness can be fit with either a S\'ersic or exponential density profile with an elliptical half-light radius $a_h$ and ellipticity $\epsilon$. The functional form of a S\'ersic is 
\begin{equation}
    \rho_{\text{S\'ersic}}(a) = \rho_0 \exp \brl -b_n \ppl \ppl \frac{a}{a_h}\ppr ^{\frac{1}{n}} -1 \ppr \brr
\end{equation}

where $b_n \approx 1.9992n - 0.3271$ and $n$ is the S\'ersic index. Assuming an elliptical system, we can solve for the normalization constant $\rho_0$ in terms of structural parameters of the galaxy to get the following density profile:

\begin{equation}
    \rho_{\text{S\'ersic}}(a) = \frac{b_n^{2n}}{2\pi a_h^2 n(1-\epsilon) \Gamma(2n) }  N_* \exp\ppl-b_n\ppl\frac{a}{a_h}\ppr^{\frac{1}{n}}\ppr
\end{equation}

Substituting n=1 into this gives the same functional form of the exponential density profile used by \cite{martin2016} and others:
\begin{equation}
    \rho_{\text{exp}}(a) = \frac{1.68^2 }  {2\pi a_h^2 (1-\epsilon)} N_* \exp(-1.68a/a_h)
\end{equation}

The elliptical radius $a$ is related to the centroid of each galaxy $(x_0, y_0)$, and the major axis position angle $\theta$ (defined to be east of north) as

\begin{multline}
a^2 = \left(\frac{1}{1-\epsilon}((x-x_0)\cos{\theta} - (y-y_0)\sin{\theta})\right)^2 \\ + \Bigl((x-x_0)\sin{\theta} + (y-y_0)\cos{\theta}\Bigr)^2
\end{multline}

We also assume a constant background density $\Sigma_b$ given by the normalized difference between the total number of observed sources in the field $N_*$ and the total number of sources in the model with an area $A$:

\begin{equation}
    \Sigma_b = \frac{N_{\text{obs}} - \int_A \rho_{\text{gal}} dA } {A}
    \label{eq:sigma_b}
\end{equation}

such that the total density profile is given by 

\begin{equation}
    \rho_{\text{model}}(a) = \rho_{\text{gal}}(a) + \Sigma_b
    \label{eq:total_density}
\end{equation}

In practice, $\Sigma_b$ takes into account spatial incompleteness by assuming that the area $A$ is not continuous and not equal to the total area of the field of view. Each ACS image has a corresponding mask of extended sources that are filtered out when performing photometry, with pixels assigned either a 0 or 1 depending on if they are contaminated or not, respectively. The total area $A$ is then given by the area per pixel of the image (from the image header) times the total number of uncontaminated pixels. \candsix and \candseven parallel fields were included in the area calculation. The integration of equation \ref{eq:sigma_b} is done numerically using a grid of 1 square arcsecond bins. 

We impose physical priors on our six free priors, all flat priors except for on $a_h$ and $n$, which have Jeffreys priors:

\begin{itemize}
    \item $|x_0| \leq 1.6$ and $|y_0| \leq 1.6$ (the center of the galaxy must be within the ACS field of view)
    \item  $\log(0) < \log(a_h) < \log(1.6')$
    \item $0 < \theta \leq \pi$
    \item $0< \epsilon \leq 1$
    \item $0< \int_A \rho_{\text{gal}} dA < N_\text{RGB}$, essentially we cannot have more than the total number of stars in our sample.
    \item $\log(0.5) \leq \log(n) < \log(8)$
\end{itemize}

\begin{figure*}
\centering
 \includegraphics[width=0.48\textwidth]{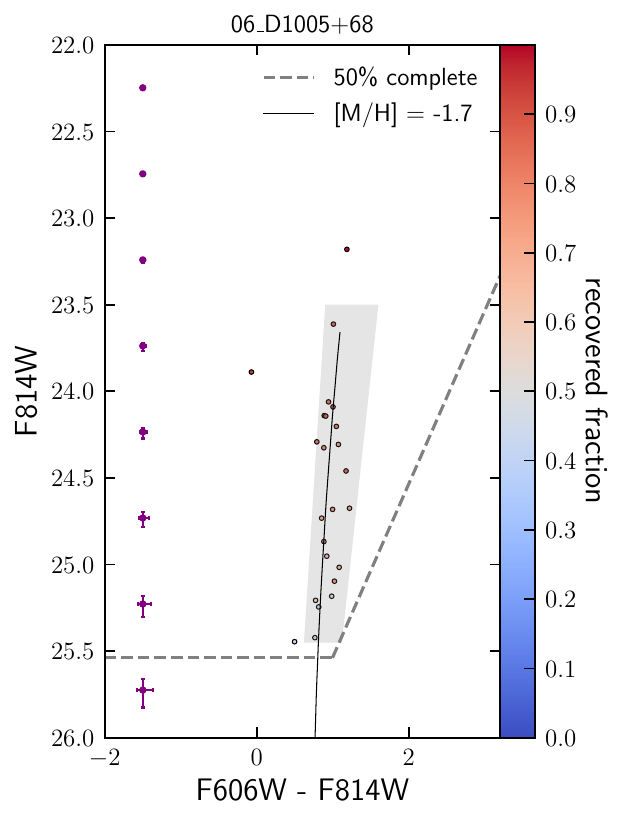}
 \includegraphics[width=0.48\textwidth]{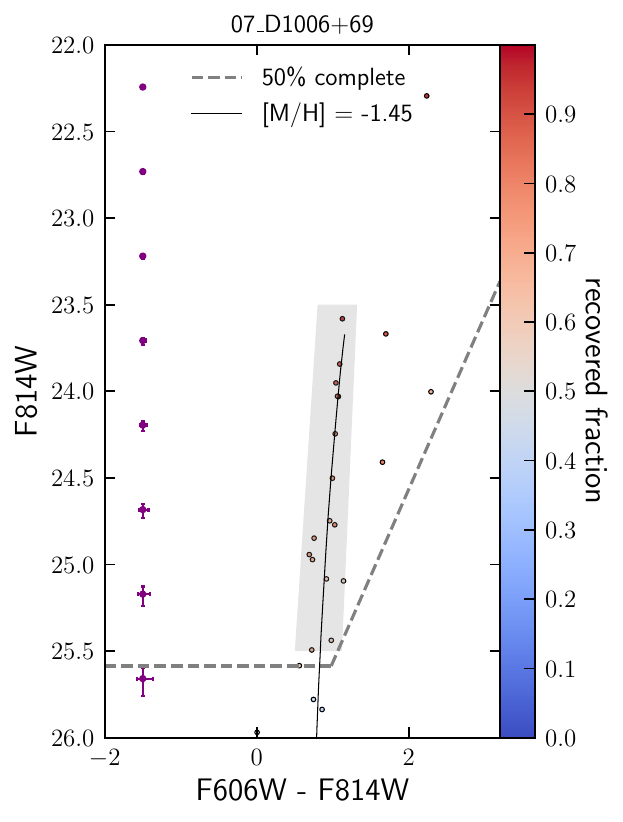}
 \includegraphics[width=0.48\textwidth]{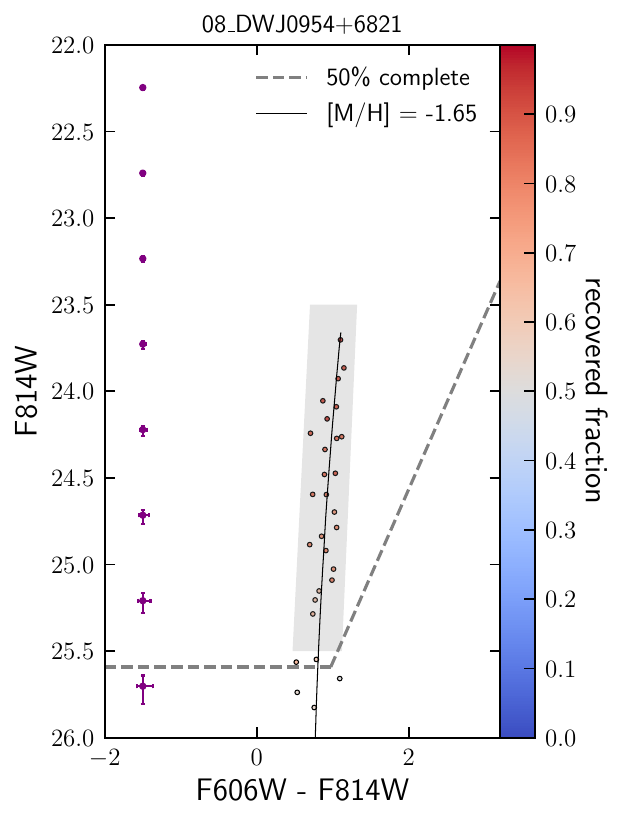}
 \includegraphics[width=0.48\textwidth]{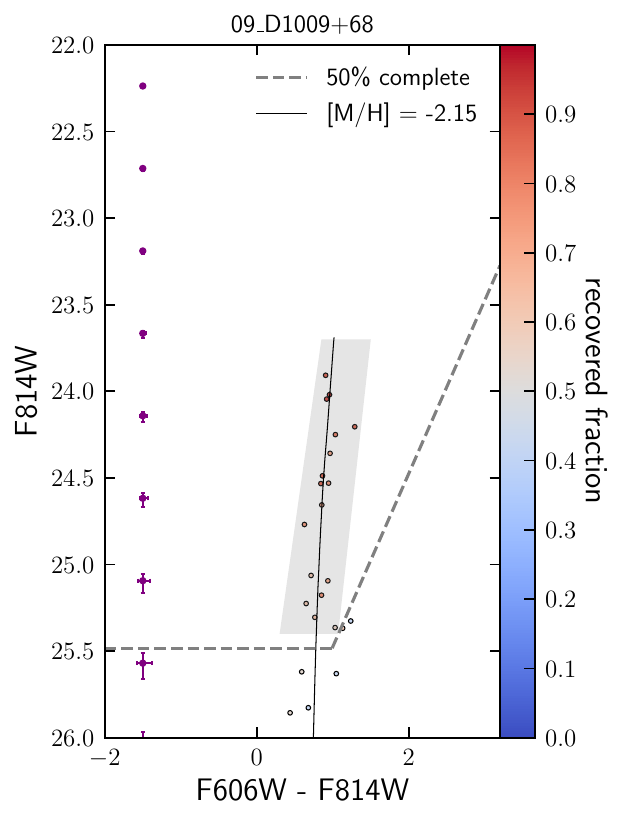}
 \caption{Color-magnitude diagrams for stars within 2$a_h$ of each satellite galaxy, as derived by the best-fit structural parameters. Points are color-coded by their completeness fraction (see Section \ref{sec:completeness}). The overall 50\% completeness limit is drawn as a dashed line. The best-fit 13 Gyr isochrone for each system is shown as a solid black line, with the metallicity derived from our RBF interpolation method. The region used to select RGB stars to use for structural parameter fitting is shown as a light gray box. Photometric errors are displayed as purple points on the left-hand side of each subplot.}
\label{fig:cmds}
\end{figure*}

We use the Python package \texttt{emcee} \citep{emcee_foremanmackey2013} to sample the posterior distribution, initialized with 32 walkers for 15,000 steps, discarding the first 2000 for burn-in. We fit only the stars selected by the RGB regions in each galaxy's CMD, as shown in Figure \ref{fig:cmds}. We report the median for each best-fit parameter and the 16th and 84th percentile values as uncertainties in Table \ref{tab:mcmc}. This table also shows other derived structural parameters and quantities of interest for each system. We note that the definition of the parameter $r_h$ varies between literature in the field; in this paper, we use $a_h$ to denote the elliptical half-light radius and $r_h$ to denote the azimuthally averaged half-light radius ($r_h = a_h \sqrt{1-\epsilon}$), as done in sources such as \cite{drlicawagner2015, richstein2024}. Note that the commonly used exponential scale length $r_e$ is related to the elliptical half-light radius via $a_h = 1.68 r_e$.

We fit all our galaxies with both exponential and S\'ersic fits and calculated the Bayseian information criterion (BIC), given as $k\log(n) - 2\log(L)$, where $k$ is the number of independent model variables (six for exponential, seven for S\'ersic), $n$ is the sample size (number of RGB stars fit), and $L$ is the maximum log likelihood of the model. A lower BIC indicates a better model fit. Only one of our satellites, D1006+69, showed a lower BIC and therefore was better fit with a S\'ersic profile: BIC$_{\text{S\'ersic}}$ - BIC$_{\text{Exp}}$ = -5, indicating the S\'ersic model is strongly preferred; therefore, we only show the exponential fits for the other three satellites.

\begin{figure*}
\centering
 \includegraphics[width=0.48\textwidth]{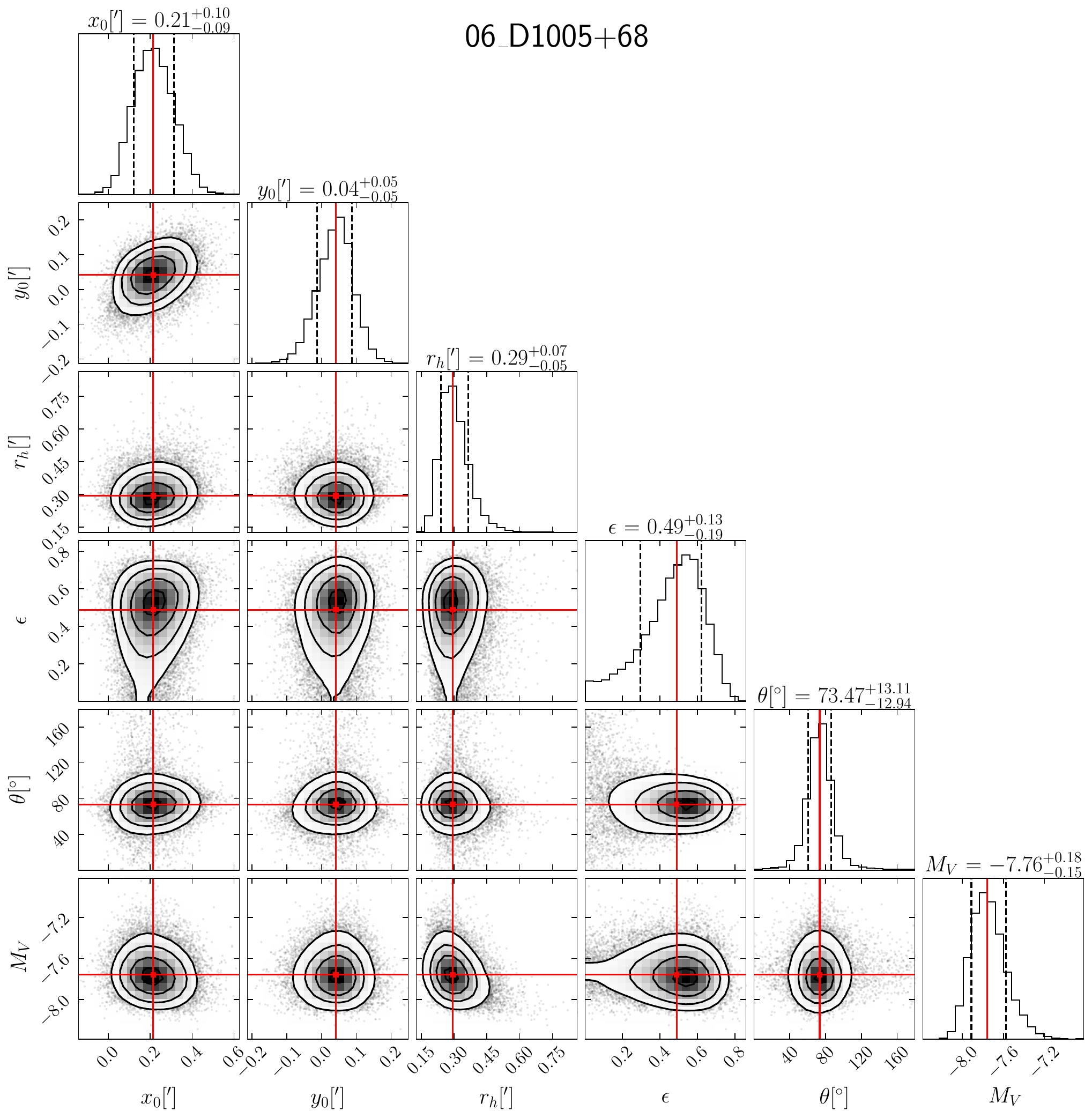}
 \includegraphics[width=0.48\textwidth]{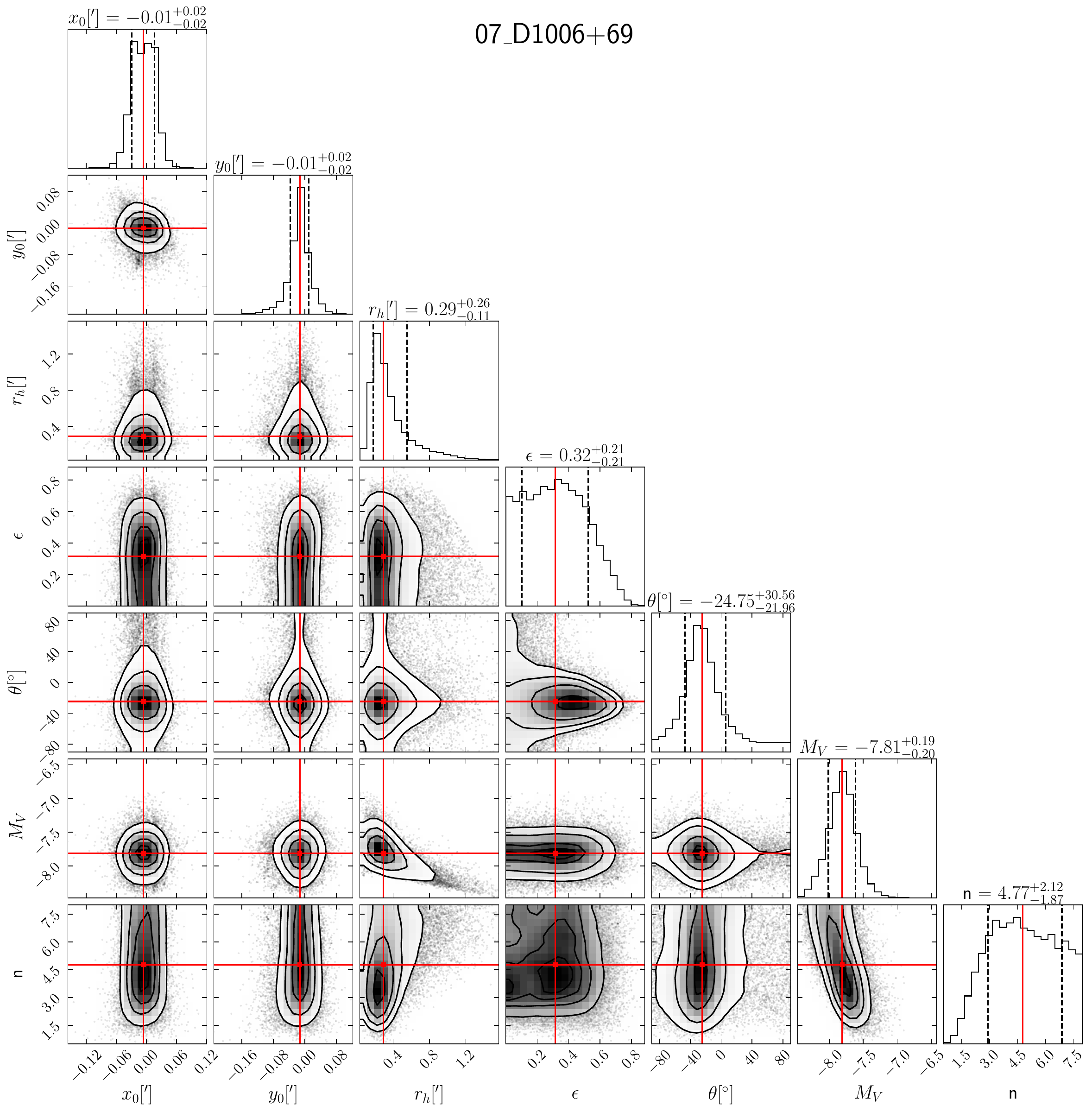}
 \includegraphics[width=0.48\textwidth]{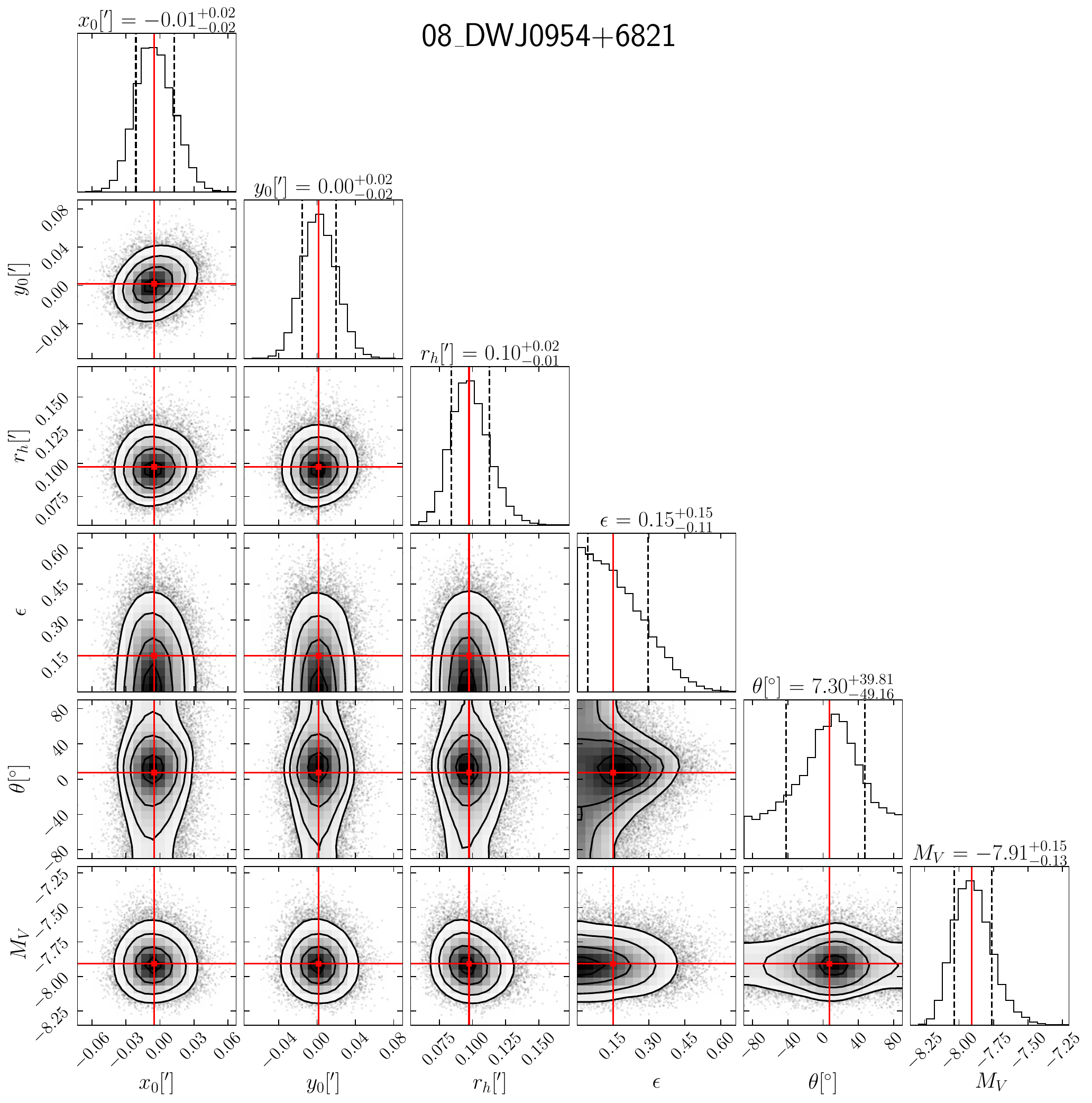}
 \includegraphics[width=0.48\textwidth]{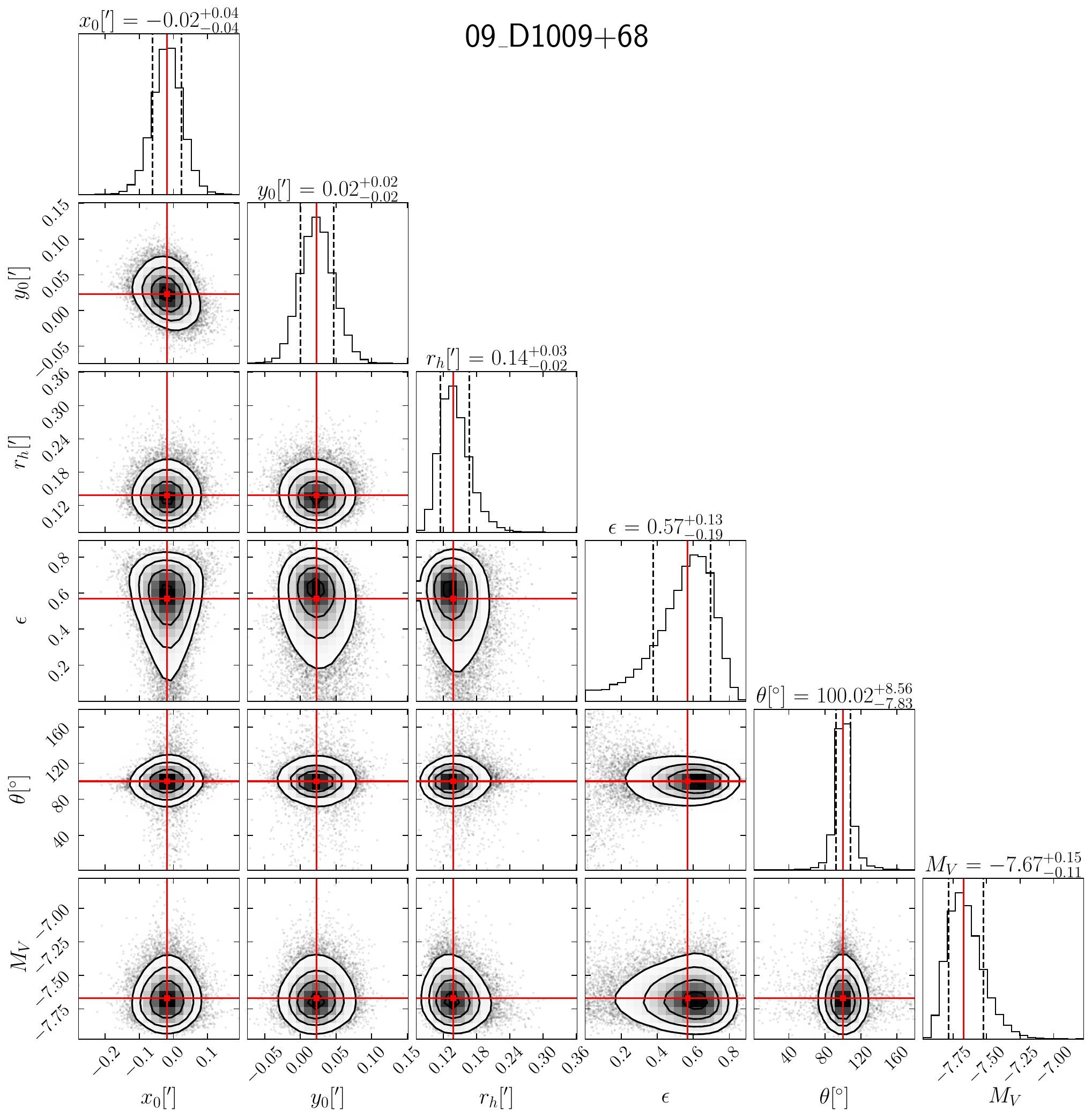}

 \caption{Corner plots showing the posterior probability distributions for various fitted structural parameters in each satellite. The circularized half-light radius $r_h = a_h \sqrt{1-\epsilon}$ is displayed instead of just the fitted elliptical half-light radius $a_h$. The median values of each parameter along with their 16th and 84th percentile uncertainties are shown as red and dashed lines, respectively. Note that while we fit for $N_*$, we instead display $M_V$, which is directly derived from $N_*$ using Equation \ref{eq:Mv} described in Section \ref{sec:absmag}. The corner plot for D1006+69 also includes the fit for its S\'ersic index.}
\label{fig:corner_plots}
\end{figure*}

\begin{figure*}
\centering
 \includegraphics[width=0.48\textwidth]{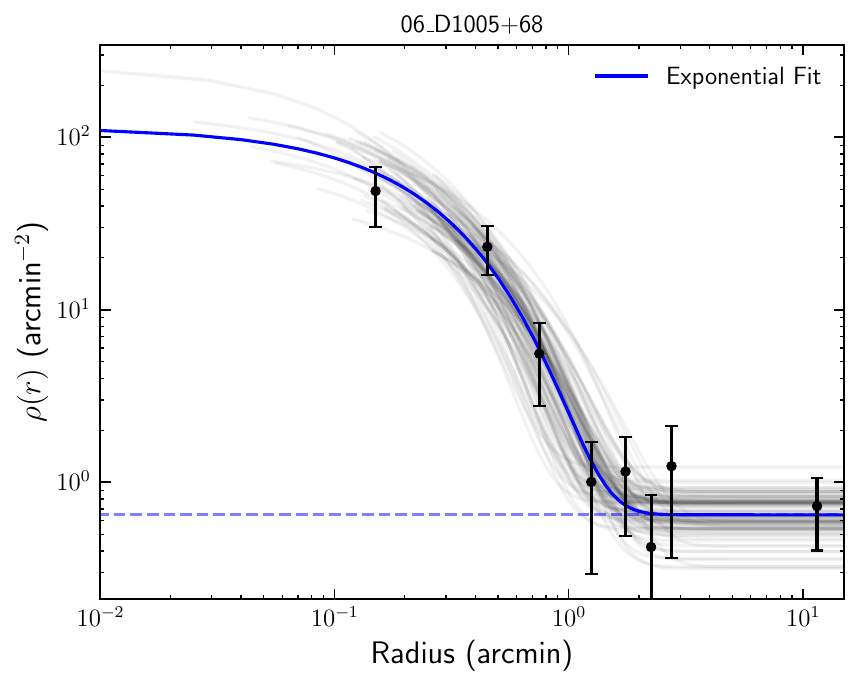}
 \includegraphics[width=0.48\textwidth]{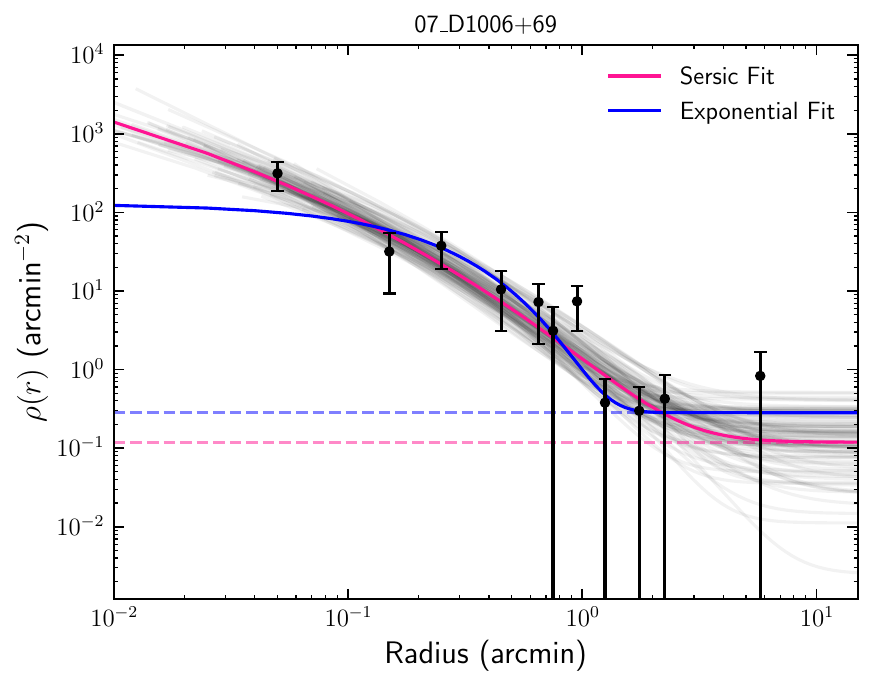}
 \includegraphics[width=0.48\textwidth]{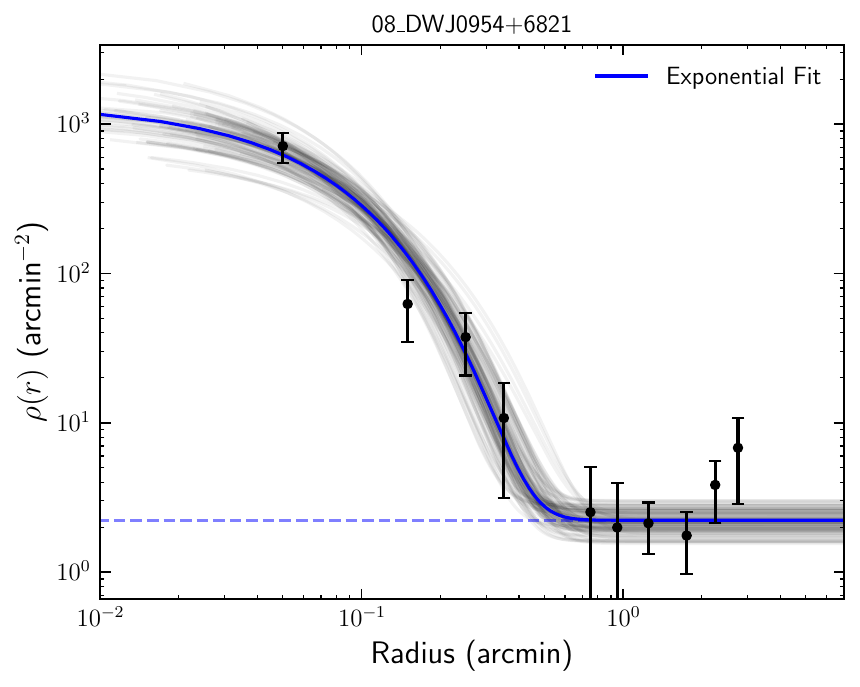}
 \includegraphics[width=0.48\textwidth]{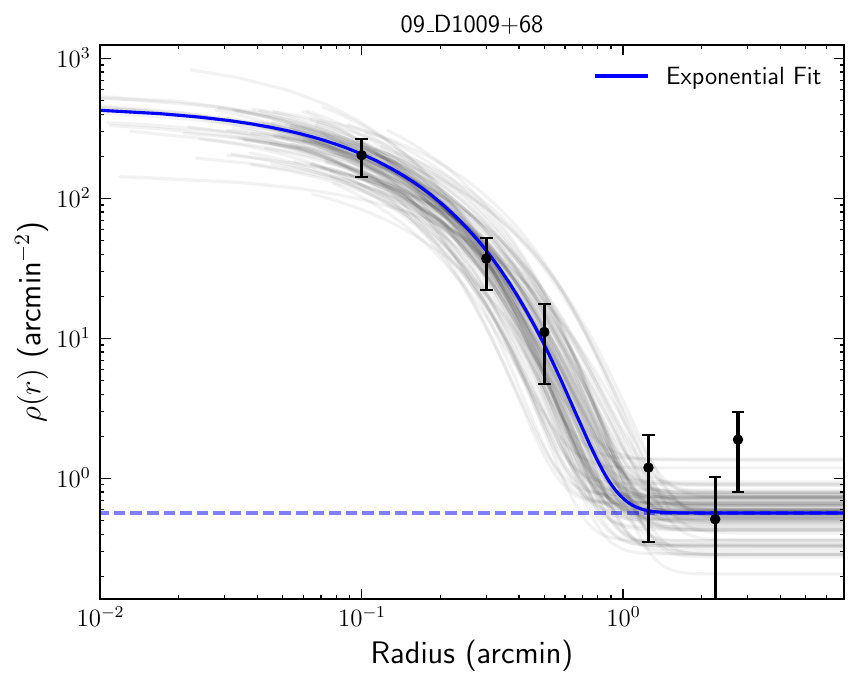}

 \caption{Radial density profiles for each satellite. Black dots are derived from counting the number of RGB stars in bins of elliptical annuli around each satellite and dividing by the total non-zero area in each bin, accounting for pixels that are masked out in each image. Error bars show Poisson uncertainties. The best-fit profile from the MCMC is shown as a colored line, with 100 random draws from the posterior distribution shown in grey.}
\label{fig:density_profiles}
\end{figure*}

\begin{figure*}
\centering

\includegraphics[width=0.49\textwidth]{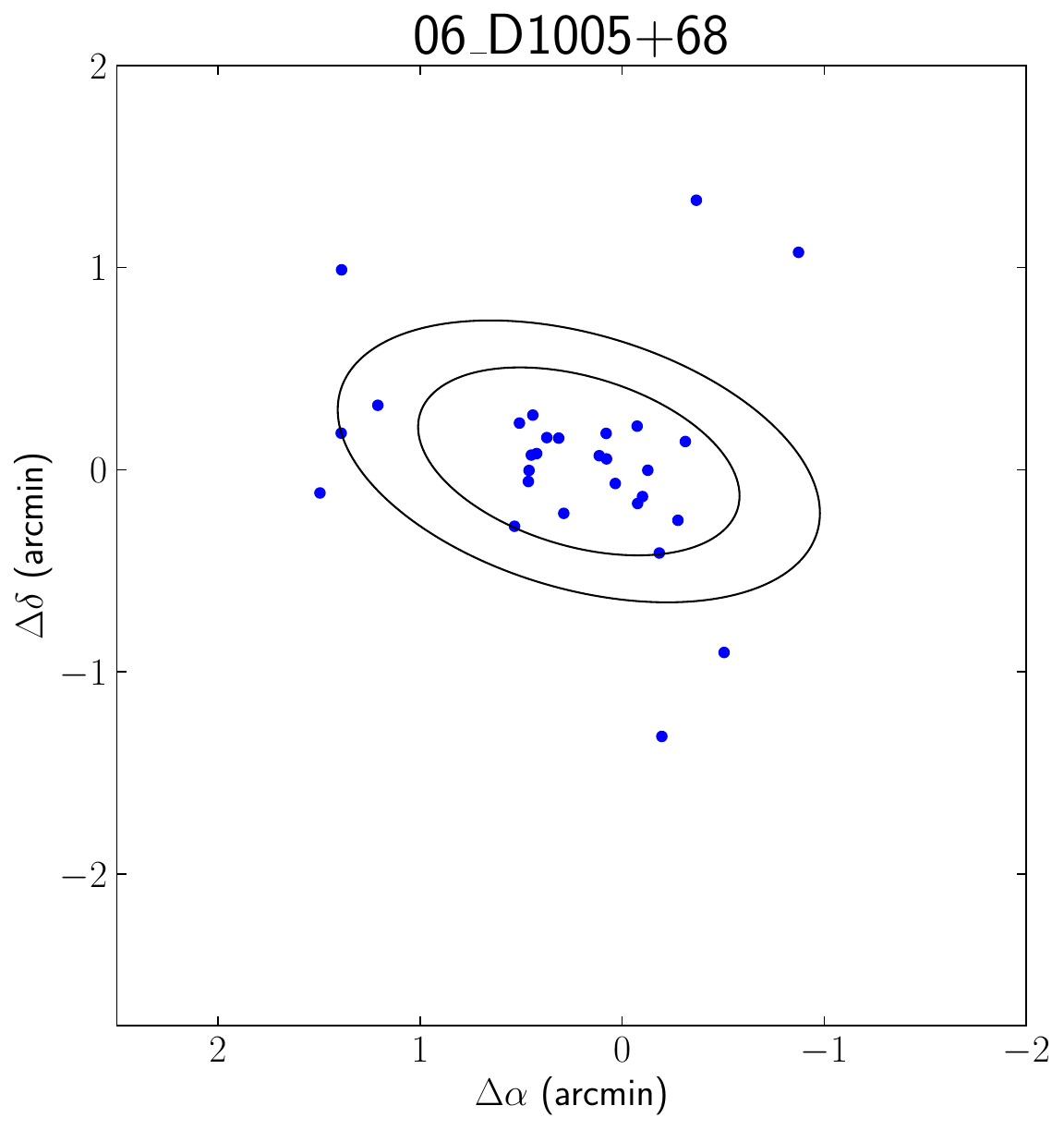}
 \includegraphics[width=0.49\textwidth]{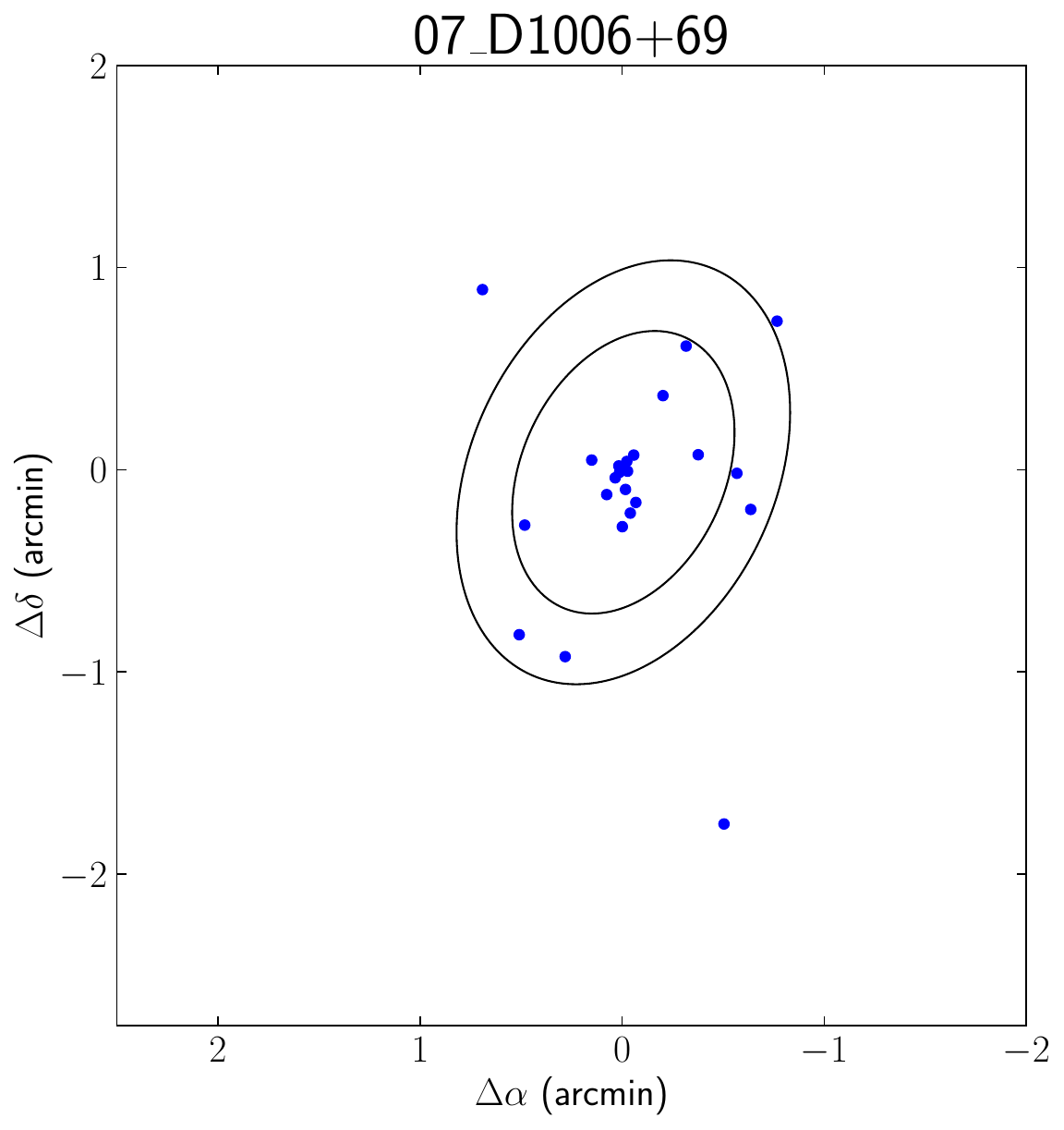}
 \includegraphics[width=0.49\textwidth]{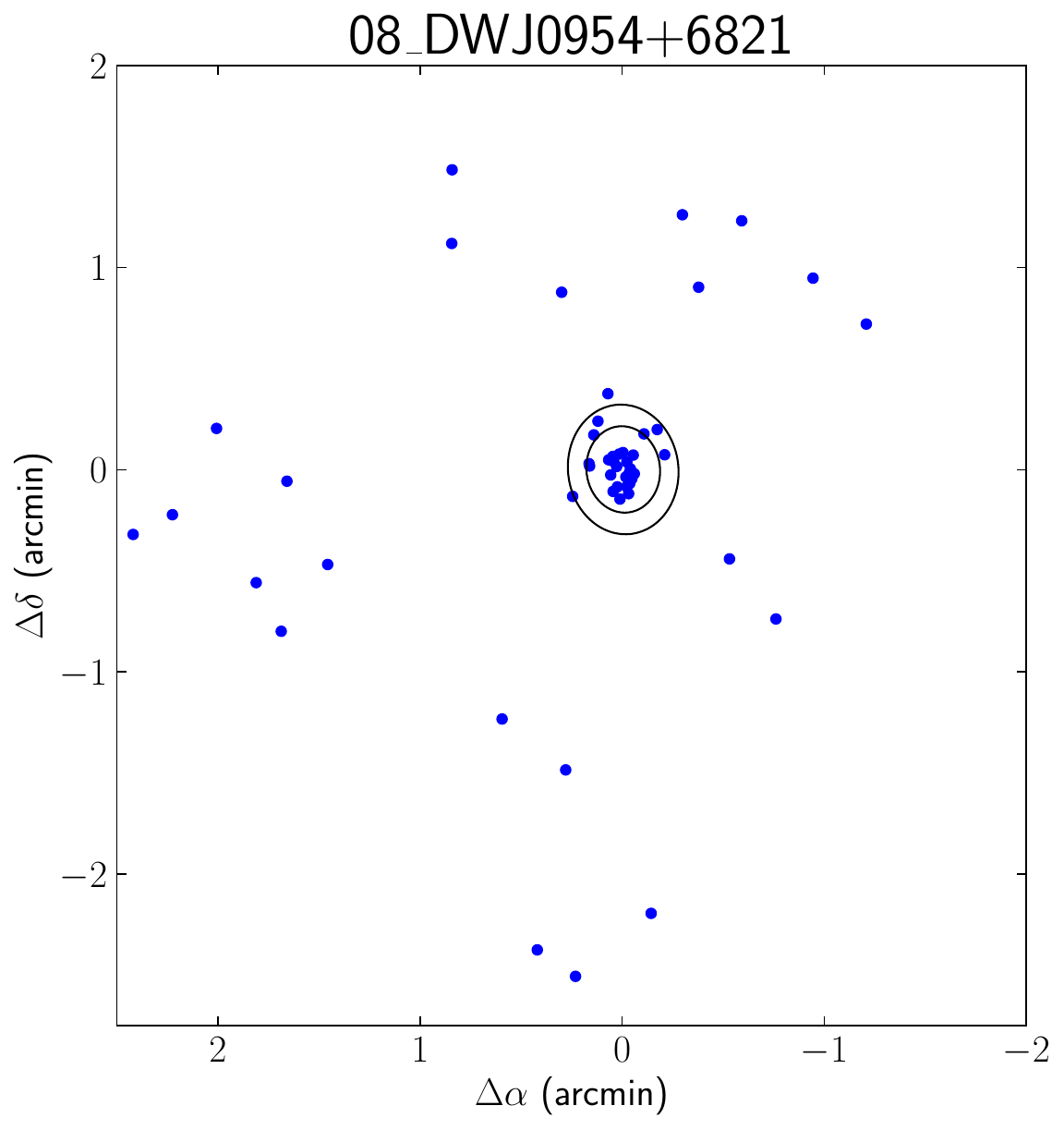}
 \includegraphics[width=0.49\textwidth]{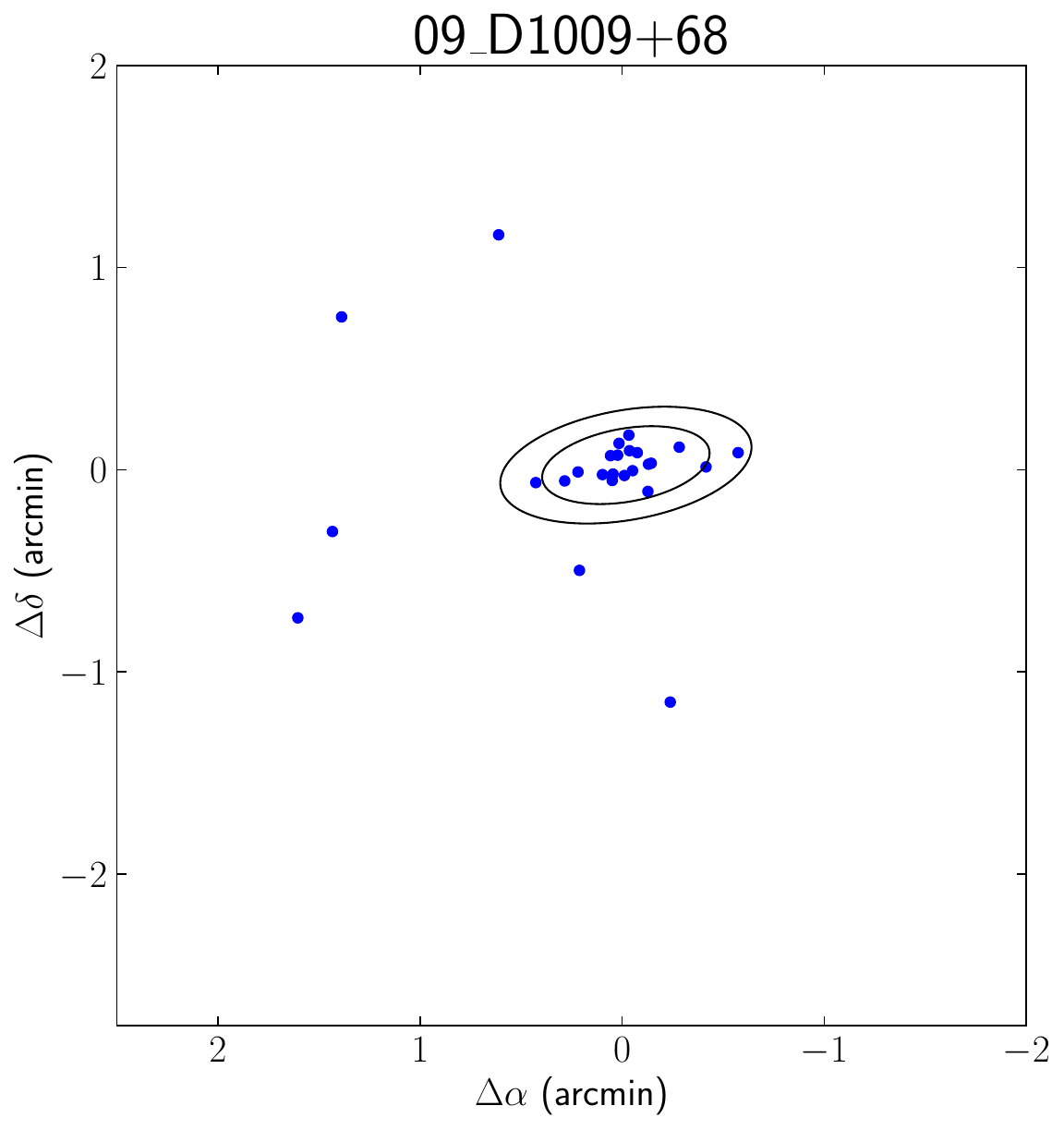}

 \caption{Distribution of point sources in each ACS field. The 2 and 3$a_h$ best-fit ellipses are shown in black. Only the RGB stars used to fit each satellite are shown. The x- and y-scales are identical in each plot.}
 \label{fig:rgb_stars}
\end{figure*}

Figure \ref{fig:corner_plots} shows the corner plots for each MCMC fit, and Figure \ref{fig:density_profiles} show the binned radial density profiles for each candidate. The best fit profile is highlighted in color and 100 random draws from the MCMC chains are shown in grey. For D1006+69, we show only the S\'ersic fit corner plot, but display both the exponential and S\'ersic density profiles, showing that even by eye, the latter profile is a better fit. We display the spatial distribution of all the RGB stars and best-fit ellipses encompassing  2 and 3$a_h$ of each satellite in Figure \ref{fig:rgb_stars}. The 2$a_h$ ellipse is also overplotted on the ACS F814W images in Figure \ref{fig:m81_map}.

Due to the small number of stars, the posterior distributions of ellipticity are not always well-constrained. Therefore, we reran our MCMC using a density profile with zero ellipticity for all four of our galaxies, and calculated the $\Delta$BIC between the elliptical and non-elliptical model fits. Based on these BIC scores, we strongly encourage the reader to think of the stated ellipticites of \candseven and \candeight as upper limits. The 84th(95th) percentile upper limits on ellipticity are 0.53(0.64) and 0.30(0.39) for \candseven and J0954+6821, respectively. We choose to use the elliptical models for the rest of our parameter derivations including $\rm M_V$ and mass, as this lets us marginalize over ellipticity as an additional source of error. These parameters do not change within error for the two different models.

%\candseven and \candeight have ellipticities of $0.32^{+0.21}_{-0.21}$ and $0.15^{+0.15}_{-0.11}$, respectively, but their posterior distributions peak near $\epsilon = 0$ and are not well-constrained Therefore, we strongly encourage the reader to think of them as upper limits. The 84th(95th) percentile upper limits on ellipticity are 0.53(0.64) and 0.30(0.39) for \candseven and J0954+6821, respectively.  As expected, D1009+68 highly ($\Delta$BIC $>3.5$) preferred an elliptical model, while \candeight extremely preferred a circular model ($\Delta$BIC $\sim -7.2$). The \candsix models were equally favored ($\Delta$BIC $\sim 0$), while D1006+69 favored a S\`ersic circular fit, but it extremely ($\Delta$BIC $\sim 19$) preferred an elliptical exponential to a circular exponential fit. 

\subsection{Photometric Metallicity}\label{sec:metallicity}
The metallicities of each satellite are derived using the method outlined in \cite{ogami2024_metallicity_rbf}. We construct a color-magnitude map using PARSEC isochrones between -2.15 $\leq$ [M/H] $\leq$ 0.5 in 0.5 dex intervals, assuming an age of 13 Gyr, a distance of 3.6 Mpc, and using only the RGB stage of each isochrone. We then interpolate these isochrones using the radial basis function (RBF) in Python's \texttt{SciPy} package, clipping values that fall outside of the interpolation range. Each star's metallicity is then the interpolated [M/H] value at the specific position of (F606W-F814W, F814W) on the CMD. We report the median metallicity of the RGB stars within 2$a_h$ as the metallicity of each galaxy. Random uncertainties are calculated by bootstrapping with replacement and finding the 16th and 84th percentiles of the bootstrapped distribution. We also calculate errors imparted by the uncertainty in distance and age of the stellar population by creating the interpolation function and evaluating it for isochrones 400 kpc closer or farther than the nominal distance of 3600 kpc, and again for isochrones of ages 8, 10, and 13 (nominal) Gyr. These are cited as separate uncertainties on the metallicity in Table \ref{tab:mcmc}. In some cases, a star in our system can fall outside of the interpolation range; this is most prominent for D1009+68, whose best-fit metallicity is the lower metallicity limit of our interpolation grid. We show an exploration of isochrones of varying age, distance, and metallicity in Appendix \ref{appendix}. Within error, our interpolation method agrees with the empirical method of \cite{streich}, who use globular-cluster-calibrated RGB color to derive metallicities for HST/ACS photometry.

In Figure \ref{fig:cmds}, we show the CMD of all point sources within 2 times the best-fit $a_h$, which we consider to be stars in each system. Each point is colored by its completeness, and a dashed line is drawn to show the 50\% completeness limit for each galaxy. The RGB box chosen is highlighted in gray. A 13 Gyr isochrone representing the median metallicity of all RGB stars is overplotted as a solid line. None of our systems possess any blue stars that could be a sign of star formation, and all are metal-poor with metallicities around [M/H] $\sim -1.5$---noting that UFDs have typical $\alpha$-enhancement of around 0.3 dex \citep{simonreview}, which is typical for UFDs around MW-mass systems)---leading us to conclude that these four systems are all likely quenched. 

\subsection{Absolute Magnitude and Mass}\label{sec:absmag}
As none of our targets show distinct TRGBs and we do not have horizontal branch stars to measure distance, we assume a distance of 3.6 Mpc for all satellites. 

The absolute magnitude of each system was derived as follows: we generated a $3.1 \times 10^6 M_\odot$ mock stellar population of age 13 Gyr and Z=0.00045 ([M/H] $\sim -1.5$) using STEV CMD 3.7 \footnote{\url{http://stev.oapd.inaf.it/cmd}} input PARSEC \citep{bressan2012_parsec} isochrones and correcting for photometric errors and completeness. From this mock sample, we pick stars corresponding to our RGB selection boxes we made for each real candidate. The magnitude of our real system should then be proportional to the magnitude of our mock stellar system: 

\begin{equation} \label{eq:Mv}
M_V = -2.5 \log_{10} \ppl {\frac{f_{mock} * N_** C_{RGB}}{N_{mock}}} \ppr
\end{equation}

where $f_{mock}$ is the total flux of the mock population, $N_*$ is the number of stars in each satellite as given by our MCMC fits, $C_{RGB} = \frac{\sum_i{1/C_i}}{N_{\text{RGB}}}$ is the mean of the inverse completeness of RGB stars for a particular satellite, and $N_{mock}$ is the number of mock stars that are within the RGB selection region of each galaxy.

A similar formula is used to calculate the mass of each system, except instead of the ratio between number of stars and flux we use the ratio between number of stars and the total mass of the mock system:

\begin{equation}
    M_{UFD} = {\frac{M_{mock} * N_* * C_{RGB}}{N_{mock}}}
\end{equation}

In our case, we query a $6 \times 10^6 M_\odot$ initial stellar mass population, which gives us a mock population with a present-day stellar mass of $2.206 \times 10^6 M_\odot$, with a fraction of the initial mass in gas and stellar remnants. Assuming a Bruzual and Charlot stellar population \citep{bruzualcharlot2003}, there is $0.152 M_\odot$ in stellar remnants compared to $0.327 M_\odot$ in stars, which suggests that the total present-day mass of our mock system is $(2.206*\frac{0.152}{0.327} + 2.206) \times 10^6 M_\odot  = 3.2 \times 10^6 M_\odot$.

The calculated $M_V$ and mass estimates are listed in Table \ref{tab:mcmc}. The uncertainty on the mass and magnitude is dictated by the uncertainty on $N_*$. Since we assume a distance of 3.6 Mpc for each satellite, we also calculate the difference in $M_V$ and mass for each satellite assuming the galaxy is at a distance of $400$ pc further or closer than M81 and also assume a 30\% error on the age of the system, following \cite{harmsen2017}. We caution that recent findings from high-resolution zoom-in simulations indicate that $M_V$ calculations from magnitude-limited data are incorrectly estimated by up to 0.3 mag for brighter dwarfs due to oversimplistic density models (see Section \ref{sec:discussion} for further discussion; \citealt{andersson2024edgeinfernosimulatingobservablestar}).

\section{Discussion}\label{sec:discussion}

\subsection{The Satellite Population of M81}
\begin{figure*}
    \centering
    \includegraphics[width=\textwidth]{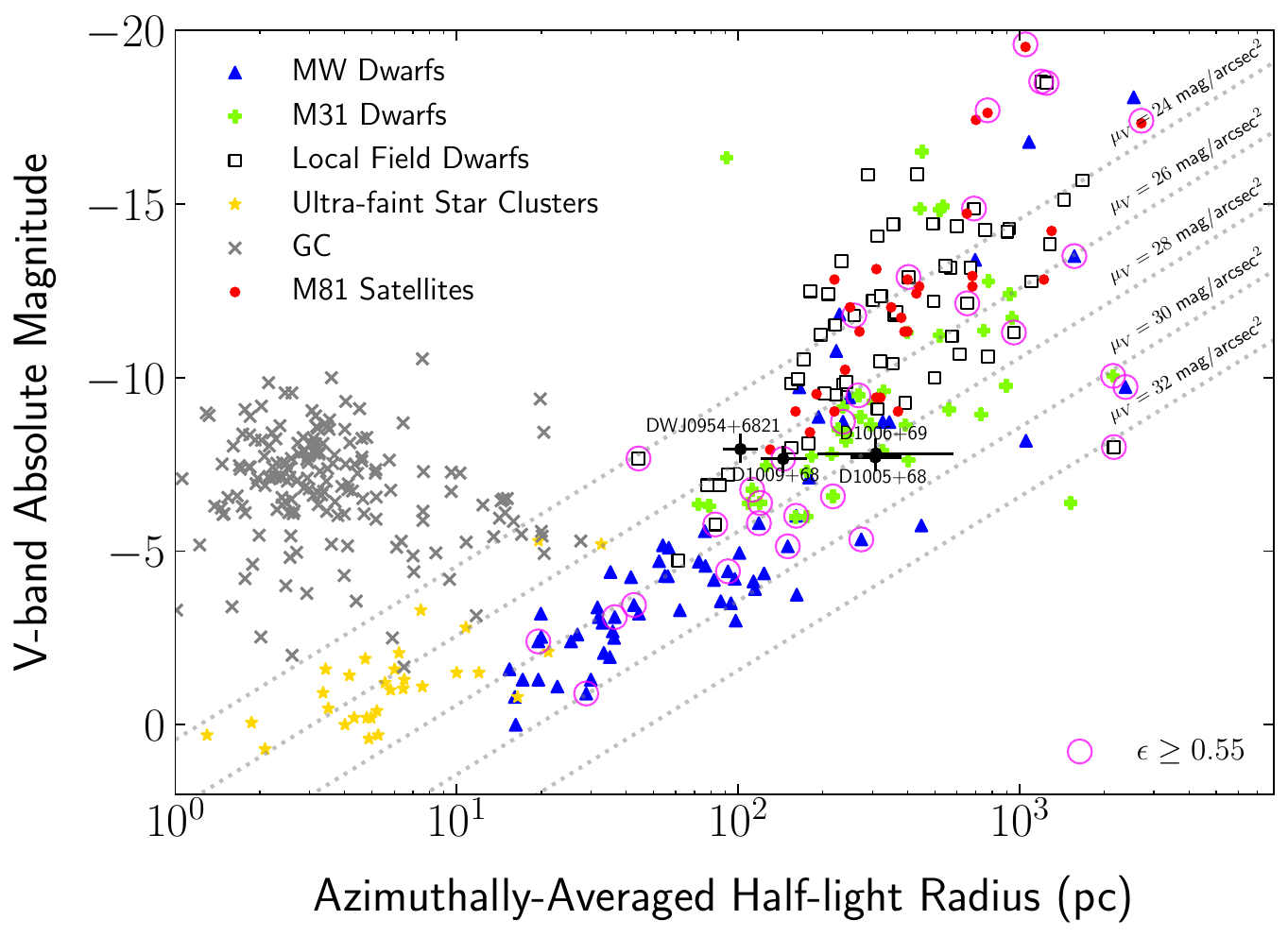}
    \caption{Distribution of Local Volume dwarf galaxies and globular clusters in the $M_V-r_h$ plane. Grey x's are globular clusters, red circles are previously known M81 satellites not characterized in this paper from \cite{chiboucas2013}. $r^\prime$ magnitudes were converted to approximate V-band magnitudes using the formula from \cite{chiboucas2009_094469}: $r^\prime = V - 0.84(\vr) + 0.13$ \citep{fukugita1996} assuming $(\vr) \sim 0.6$. All other colored points are from the Local Volume Database. The four galaxies highlighted in this work are shown as black circles with errorbars. The conversion for $r_h$ from arcmin to parsecs for these galaxies is done assuming the distance of M81, $D=3.6$ Mpc. Points outlined in a pink oval are galaxies that are substantially elliptical ($\epsilon\geq0.55$), which includes D1009+68 from this work, making it one of the most elliptical UFDs ever found. 
    %{\color{red} "GC" could be "GCs"}
    }
    \label{fig:radiuslum}
\end{figure*}

The four dwarf galaxies investigated in this work exemplify the diversity of satellite populations in MW-mass galaxies. We place them in context with other known satellites around the MW, M31, and M81 in Figure \ref{fig:radiuslum}, where we show the V-band absolute magnitude against the azimuthally-averaged half light radius of each stellar system. Though they are of similar magnitude, our four systems span an order of magnitude in sizes and are amongst the faintest M81 companions known. \candeight is the most compact M81 satellite found, with a circular half-light radius of $ 0.10^{+0.02}_{-0.01}$ arcmin or $102_{-17}^{+21}$ pc across. While our satellites have varied structural properties, none of them show evidence of recent star formation. Despite the M81 group's more active satellite star-formation history, its faintest neighbors are all uniformly quenched.

\begin{figure*}
    \centering
    \includegraphics[width=\textwidth]{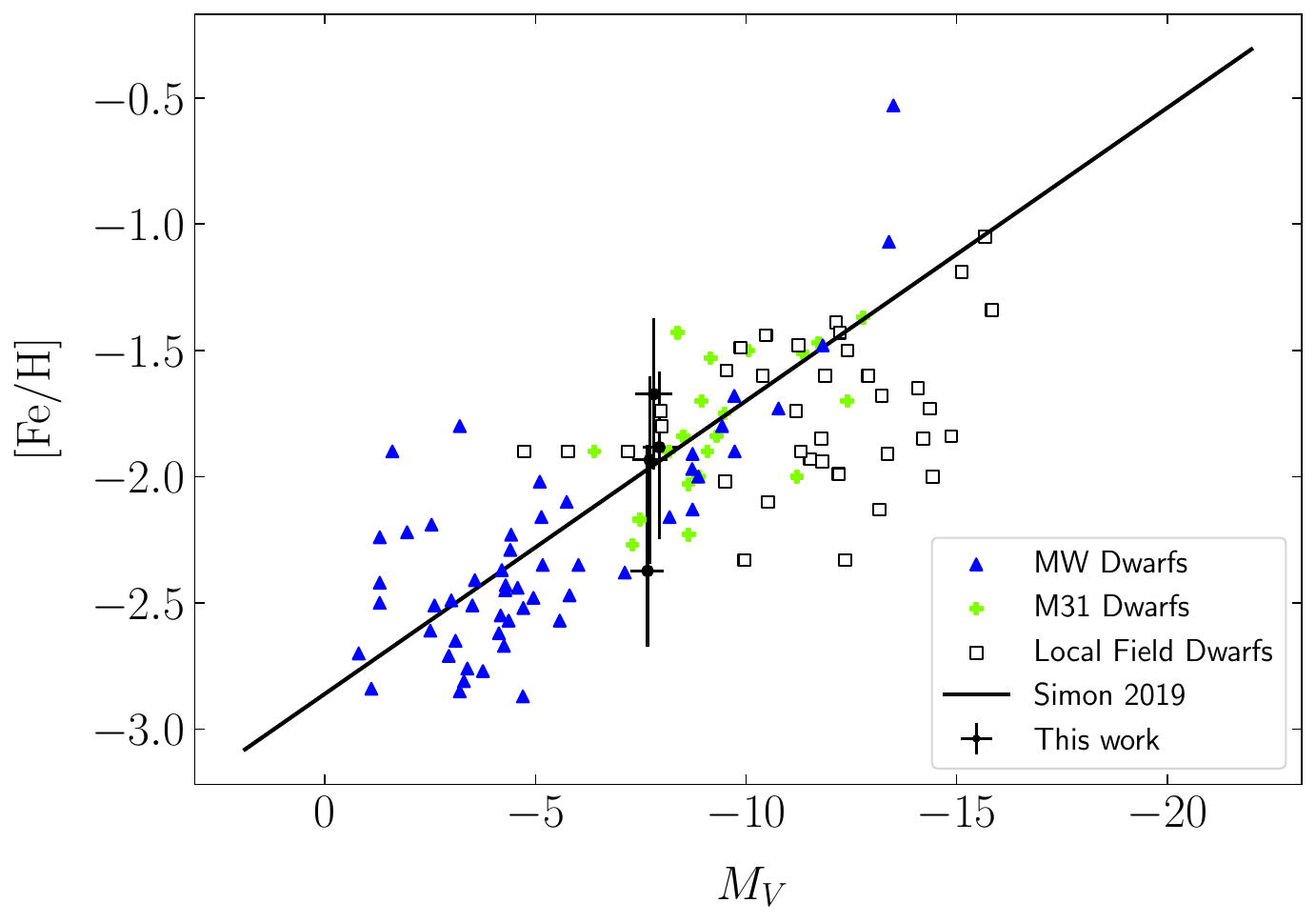}
    \caption{Luminosity-metallicity relation for nearby satellites. Metallicities of satellites from this work are converted to [Fe/H] using \citealt{streich} assuming an $[\alpha/ \rm Fe] = 0.3$. All other data points are from the Local Volume Database. Error bars include errors from bootstrapping and errors on age. If the combined mean metallicity error for a particular galaxy was less than the average combined mean error of all four galaxies, we set the error to 0.3 dex.}
    \label{fig:massmet}
\end{figure*}

\cite{kirby2013_massmet} show that dwarf satellites of the MW and M31 obey an identical stellar mass-metallicity relation that is an extension of the relation found in more massive galaxies, with more scatter at the low-mass end. We plot this relation as shown in Figure \ref{fig:massmet}, using V-band luminosity as a proxy for stellar mass. Since we measure [M/H] from our isochrones, we convert to an [Fe/H] estimate based on \cite{streich} assuming an $[\alpha/\text{Fe}] = 0.3$, as is typical in UFD stars \citep{simonreview}. Error bars on metallicity include random error from bootstrapping and age but without additional uncertainty from distance. In cases where the total uncertainty for a galaxy is less than the combined mean uncertainty for all four galaxies, we chose to set its metallicity uncertainty to 0.3 dex. MW and M31 satellite metallicites are spectroscopic unless only photometric measures are available. All values other than those from this work are from the Local Volume Database \citep{pace2024_local_volume_database}\footnote{\url{https://github.com/apace7/local_volume_database }}. The black line shows the updated best-fit from \cite{simonreview}, consistent with \citet{kirby2013_massmet}. We find that all four of the satellites analysed in this work are consistent with the luminosity-metallicity relation within observational scatter, which would be expected if these systems were indeed low-luminosity galaxies and not star clusters. Based on their properties, we confirm these systems as ultra-faint and near-ultra-faint dwarf galaxies in the M81 system.

\subsection{Tidal Radii and Ellipticities}

Out of all the galaxies in our sample, \candnine is by far the most elliptical, with $\epsilon = 0.57_{-0.19}^{+0.13}$, while \candeight is the least, with $\epsilon < 0.30$. One may expect that satellites closer to their host are more likely to be tidally interacting and therefore have a larger ellipticity \citep{martin2008, longeard2022_bootesI}. We investigate this by estimating the tidal radii of our candidates. We assume that $r_{\text{tidal}} \sim R \ppl \frac{M_{sat}}{2M_{enc}(R)}\ppr^{1/3}$, where R is the separation between the central galaxy and satellite and $M_{enc}(R) = \frac{v_c^2 R}{G}$, where $v_c = 230$ km/s as given by M81's H1 rotation curve \citep{deblok2008_m81rotationcurve}. Since our galaxies are extremely faint, it is likely that they are dark-matter dominated, therefore $M_{\text{tot}} \gg M_{*}$. Though we do not know the total mass (including dark matter) of each satellite, we estimate the satellite mass using its dynamical mass as derived in \cite{wolf2010_dynamicalmass}, assuming dispersion-dominated systems: 
\begin{equation}
    M_{1/2} \simeq 930 \ppl \frac{\sigma_{\text{los}}^2} {\text{km}^2 \text{s}^{-2}}\ppr \ppl \frac{r_{\text{h,phys}}}{\text{pc}} \ppr M_\odot.
\end{equation}

We estimate $\sigma_{\text{los}} \sim 5$ km/s as a characteristic line-of-sight velocity dispersion for our systems, a typical dispersion recorded for UFDs in MW-like systems.
The tidal radii calculated are listed in Table \ref{tab:mcmc}. We note that the total mass of each satellite is greater than the estimated dynamical mass, making our tidal radii underestimated. Nonetheless, the current estimated tidal radii are much greater than twice the half-light radius of each system, indicating that none of our systems are likely being stripped or affected by tides at their outskirts. Repeating this exercise assuming tidal interaction with NGC 3077 instead of M82, using a rotation velocity of $v_c = 65$ km/s \citep{martin1998ApJ_ngc3077rotation}, yields the same conclusion, with the $R_{\text{tidal}}$ even larger than assuming M81's potential.

But how do we reconcile this with our ellipticity measurements?  Both simulations and observations point to there being little correlation between the ellipticity of satellites and their distances from their host galaxy.

\cite{goater2024_EDGE_stellarhalos} simulates void galaxies in the EDGE simulation and finds that, despite being tidally isolated, their galaxies still possessed the wide range of ellipticities seen in observed UFDs. They argue that even with changes to feedback prescription, the elongated structures of these simulated dwarfs can arise due to a dry merger at late times rather than being a tidal tail, and that more elliptical UFDs have later formation times and lower surface brightnesses. 

\begin{figure*}
    \centering
    \includegraphics[width=\textwidth]{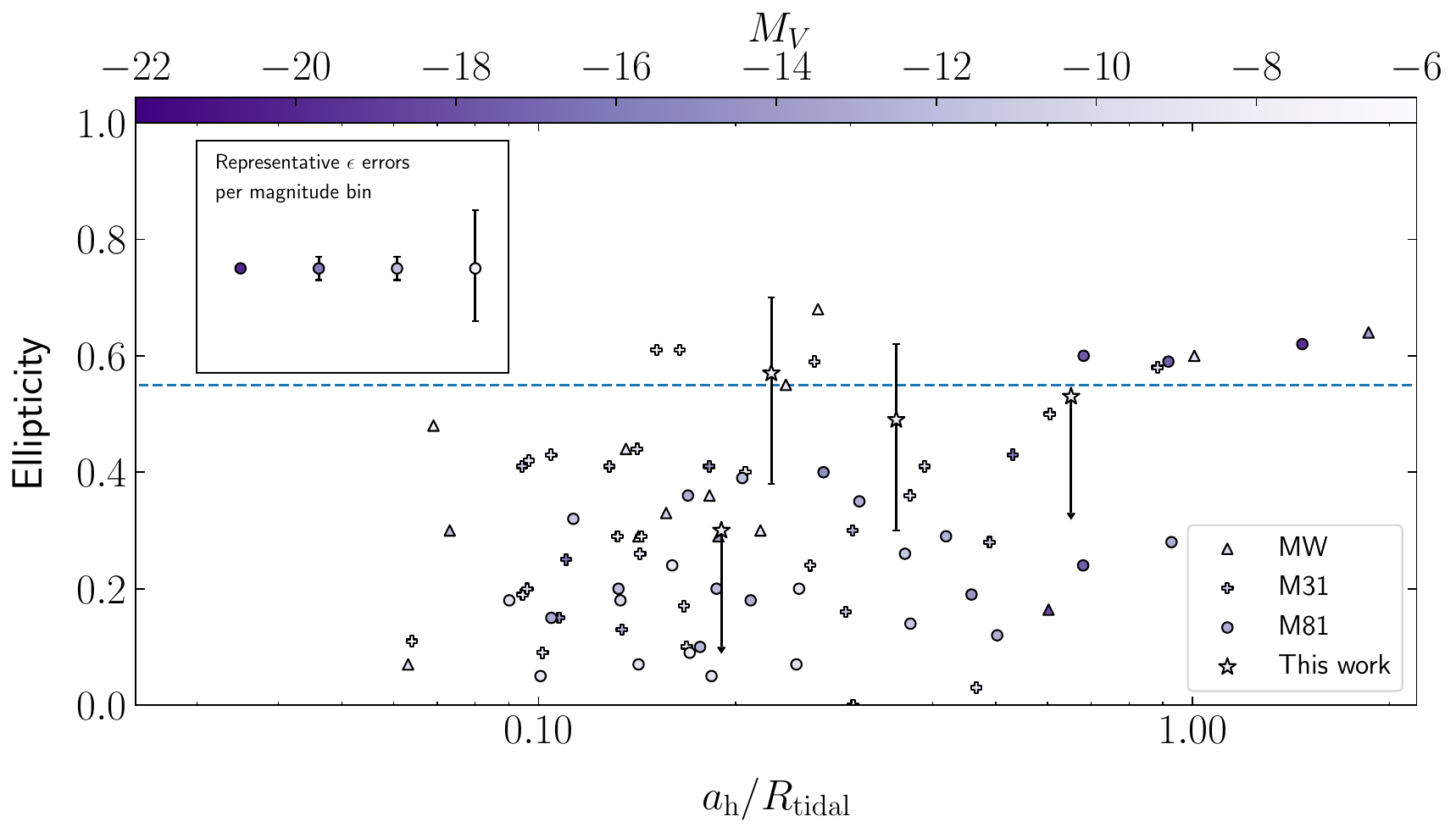}
    \caption{The ellipticity of select Local Volume satellites against the ratio of their elliptical half light radii and tidal radii. Tidal radii are calculated assuming a systemic line-of-sight velocity dispersion of 5 km/s for M81 satellites and a rotational velocity of 230 km/s for all hosts. Each point is color-coded by the satellite's absolute magnitude, $M_V$. We plot only  well-measured satellites in the restricted luminosity range of $M_V < -6$. \candseven and \candeight ellipticities are shown as 84\% confidence upper limits. A blue horizontal line is drawn at an ellipticity of 0.55, matching the ellipticity cut in Figure \ref{fig:radiuslum} for highly elliptical satellites. Median errors in ellipticity are shown in the inset box for various magnitudes as binned from $M_V = -22$ to $M_V = -6$ in 4 dex bins. These errors are only calculated from the MW and M31 satellites, as \cite{chiboucas2009_094469} do not cite error bars for their ellipticity measurements. We do not see any trends between ellipticity and tidal radius fraction, indicating that tides are likely not the dominant factor that sets the ellipticity of low-mass satellites. Data for MW and M31 satellites are from the Local Volume Database, data for M81 satellites are compiled from \cite{chiboucas2009_094469}.}
    \label{fig:ell_vs_tidal}
\end{figure*}

Observations of dwarf satellites also paint the picture that these systems can have a diverse range of ellipticities which are not correlated with other physical properties. Figure \ref{fig:ell_vs_tidal} shows the ellipticity of well-measured ($\rm M_V<-6$) MW, M31, and M81 satellites as a function of the ratio of their elliptical half-light radii and tidal radii. The tidal radii for MW, M31, and other M81 satellites are calculated the same way as for our satellites. We assume a $v_c = 230$ km/s for all host systems. MW and M31 satellites have listed systemic line-of-sight velocity dispersions in the Local Volume Database, while for M81 we assume $\sigma_{\rm los} = 5$ km/s. These points are colored by the absolute V-band magnitude of the satellite as reported in the literature. We see no trends in ellipticity as a function of either tidal radius fraction or absolute magnitude --- satellites that are close in size to their tidal radii are not necessarily more elongated or fainter, but instead span the full gamut of ellipticites and luminosities that have been measured for satellite dwarfs. In fact, Hercules, the most elliptical satellite in this sample with an $\epsilon = 0.69 \pm 0.3$ has a smaller tidal ratio than any of the four satellites in this work, while Sagittarius has the highest at almost 2, despite having almost an identical ellipticity as Hercules. No trend is seen in ellipticity as a function of separation from the central galaxy either (see Figure \ref{fig:ell_vs_sep} in Appendix \ref{ellvsep}). Cetus III, the most elliptical satellite with $\epsilon \sim 0.76^{+0.08}_{-0.06}$ is one of the more distant satellites, with a separation of $\sim 250$ kpc. This indicates that tidal stripping is likely not what is setting the ellipticity of many satellites and, indeed, satellites around MW-mass hosts can exhibit a wide range of ellipticities at any separation. 

\subsection{D1006+69: The Case of a High S\'ersic Index}
The S\'ersic index $n$ is a measure of a galaxy's concentration. A higher $n$ means a more centrally-concentrated galaxy and generally correlates with a galaxy that is larger in radius and brighter. Elliptical galaxies generally have $n \sim 4$, while disks are generally better fit with $n \sim 1$ \citep{TRUJILLO2001_sersic}. Interestingly, \cite{marchilasch2019_sersic_vs_reff} find an anticorrelation between the S\'ersic index and the effective radius in a sample of dwarf and UFD galaxies, making \candseven an outlier in this relation as we find that it is large and also very centrally concentrated. 

Resolved-star UFD fitting has historically been done using an exponential ($n=1$) fit with great accuracy, following \cite{martin2008}. \cite{munoz2018_satfit_sersic} fit satellites with a wide range of luminosities using S\'ersic, Plummer, and exponential profiles, and found all of the UFDs in their sample were best fit with low, $n \sim 1$ S\'ersic indices. That said, there is an outlier: \cite{chiti2021NatAs_tucII} identified that the Tucana II dwarf is much more extended than originally thought and find that it has a spatially extended stellar halo. \cite{tarumi2021_tucanaII} use cosmological hydrodynamic simulations to show that a galaxy with an extended distribution and a de Vaucouleurs profile ($n \sim 4$) can be created through the major merger of two low-mass galaxies. This opens up the interpretation that \candseven may host an accreted stellar halo. Other UFD systems have also been found that possess stars at unusually large radii. \cite{waller2023_RRL_outskirts} identified member stars in the outskirts of Coma Berenices, Ursa Major I, and Bo\"otes I whose chemistry is compatible with that of the central stellar populations. They propose that the former two systems underwent supernova feedback or tidal stripping, which caused the migration of stars to larger radii, while the latter system, Bo\"otes I (also severely elongated with an ellipticity $\epsilon = 0.68 \pm 0.15$;  \citealt{longeard2022_bootesI}), may have undergone a minor merger that deposited its outermost stars in a halo. Most recently, a number of papers have found evidence for extended stellar populations forming stellar halos around dwarf galaxies, with stars several half-light radii away from their centers, including \cite{chiti2021NatAs_tucII, longeard2022_bootesI, tau2024_UFD_stellarhalos} and \cite{jensen2024_UFDhalos}. Though all of these extended systems are satellites of the Milky Way, there is no reason to believe that similarly extended UFDs possessing faint stellar halos cannot exist in other MW-mass systems. Indeed, recent zoom-in cosmological simulations of faint, low-mass, isolated galaxies found that each simulated dwarf was best fit with a multicomponent light profile which included at least both a central exponential and an extended outer halo component. Using only a single exponential component leads to an underestimation in $M_V$ from observational data \citep{andersson2024edgeinfernosimulatingobservablestar}. This is not accounted for by most current UFD studies; we partially account for this by using a S\'ersic profile for D1006+69, but understanding how to best include multiple components in fitting is the subject of future work in the community.

%\subsection{\candeight}
%We confirm that \candeight is a newly discovered, confirmed UFD system in the M81 group. With an $a_h=xxx$ and an $\epsilon=XXX$, its structural parameters point to it being one of the most compact and circular satellites in the system.

\subsection{Outlook and Future Facilities}

With first light of the Rubin Observatory only a year away, we will soon have access to a very wide-field resolved-star survey, that after 5--10 years will have comparable depth to the M81 group data across most of the Southern sky. It is anticipated that Rubin will be able to complete the dwarf galaxy census down to M$_{\rm V} \sim -7$ at 3.5 Mpc \citep{mutlupakdil2021_rubin}, which matches our current analyses. Due to Rubin's location in the Southern Hemisphere, only a handful of MW-mass groups can be surveyed (e.g. NGC 253, Cen A), but it will be able to survey lower mass groups of similar distance, such as NGCs 247, 55, and 3109, many of which have only been surveyed out to a fraction of their virial radii \citep{whiting1997AJ_ngc3019antila, mcnanna2024_ngc55_sats, romanowsky2023_ngc247_faintfuzzy}. Rubin will scan the entire sky and find satellites not only at or beyond the virial radius of these sparsely surveyed systems, but also has the potential to discover field UFDs whose isolated nature makes them a gold mine for understanding UFD properties independently of environment. 

While Rubin will excel in discovering previously unattainable satellites, our work reveals an ever-present necessity for a synergy with high-resolution space-based follow-up. \candseven and \candnine were found to have absolute magnitudes of $M_V = -8.91 \pm 0.40$ and $M_V = -8.73 \pm 0.45$, respectively, by \cite{okamoto2019_d100669}, while we find them to be much fainter at $M_V \sim -7.8$ and $M_V \sim -7.7$. Our analysis of \candseven also revealed a widely different S\'ersic index than \cite{okamoto2019_d100669} -- while their analysis found $n=1.04 \pm 0.96$, ours makes it the most centrally concentrated UFD satellite ever found with $n=4.77^{+2.12}_{-1.87}$. These major differences in structural properties highlight the uncertainty of ground-based characterization and is a lesson that space-based follow-up will be highly important even in the Rubin era.

In tandem with Rubin, the Nancy Grace Roman Space Telescope will be able to probe down to 27th-28th near-IR magnitude in one hour.  With its unprecedented depth, it will be able to resolve RGB stars in galactic stellar halos out to $D \leq 10$ Mpc and lend itself to the discovery, confirmation, and probing of stellar populations in the ``bright" UFDs of $M_V \sim -7$ within galaxy groups at $D \sim 3.5$ Mpc such as M81, NGC 253, and Cen A \citep{walter2022AJ_roman, wright2023_foggie_roman}. Confirming candidates with such low surface brightnesses will require JWST NIRCam follow-up, which would enable us to probe down to horizontal branch stars to perform detailed structural analysis and derive robust distance measurements.

\section{Conclusion}\label{sec:conclusion}
In this work, we have used HST snapshot imaging to measure the structural parameters and confirm the galactic and faint nature of four satellites in the M81 group: D1005+68, D1006+69, DWJ0954+6821, and D1009+68. %Three of these satellites, D1005+68, D1006+69, and D1009+68, were previously discovered using ground-based data and one of them, DWJ0954+6821, is a new discovery based on imaging from Subaru/HSC that has been followed-up and confirmed in this study. 
The properties derived in this paper establish the heterogeneity of satellite properties of MW-mass galaxies. \candeight is the most compact and circular system found in the M81 group, \candseven is the most concentrated, and \candnine is one of the most elliptical faint dwarf galaxies found among Local Volume satellites. All four galaxies are additionally the faintest M81 satellites ever discovered. Analysis of their stellar populations did not yield any indications of recent or ongoing star-formation, and calculations of their tidal radii showed that none of these satellites are likely being tidally stripped by M81.

The upcoming Rubin Telescope and Roman Space Telescope will revolutionize our searches for nearby, extremely faint dwarf satellite galaxies. Harnessing the wide breadth of Rubin's survey strategy with the incredible depth of Roman, we will be able to measure the satellite luminosity function and characterize UFD stellar populations out to 10 Mpc, thereby putting constraints on and enabling comparisons with cosmological simulations.

\section{Acknowledgments}
We thank the anonymous referee for comments that greatly improved this manuscript. This work was partly supported by HST grant GO-17158 provided by NASA through a grant from the Space Telescope Science Institute, which is operated by the Association of Universities for Research in Astronomy, Inc., under NASA contract NAS5-26555, the National Science Foundation through grant NSF-AST 2007065, and by the WFIRST Infrared Nearby Galaxies Survey (WINGS) collaboration through NASA grant NNG16PJ28C through subcontract from the University of Washington. A.M. gratefully acknowledges support by FONDECYT Regular grant 1212046 and by the ANID BASAL project FB210003, as well as funding from the Max Planck Society through a ``PartnerGroup” grant. This research has made use of NASA's Astrophysics Data System Bibliographic Services. 

Based on observations utilizing the Pan-STARRS1 Survey. The Pan-STARRS1 Surveys (PS1) and the PS1 public science archive have been made possible through contributions by the Institute for Astronomy, the University of Hawaii, the Pan-STARRS Project Office, the Max-Planck Society and its participating institutes, the Max Planck Institute for Astronomy, Heidelberg and the Max Planck Institute for Extraterrestrial Physics, Garching, The Johns Hopkins University, Durham University, the University of Edinburgh, the Queen’s University Belfast, the Harvard-Smithsonian Center for Astrophysics, the Las Cumbres Observatory Global Telescope Network Incorporated, the National Central University of Taiwan, the Space Telescope Science Institute, the National Aeronautics and Space Administration under Grant No. NNX08AR22G issued through the Planetary Science Division of the NASA Science Mission Directorate, the National Science Foundation Grant No. AST-1238877, the University of Maryland, Eotvos Lorand University (ELTE), the Los Alamos National Laboratory, and the Gordon and Betty Moore Foundation. 
%\end{acknowledgments}

\vspace{5mm}
\facilities{HST/ACS} 

\software{\code{DOLPHOT} \citep{dolphot2000, dolphot2016}, Aladin \citep{aladin2000}, \code{Python 3.11} (\citealt{python}) \code{— Matplotlib} (\citealt{matplotlib}), \code{Numpy} (\citealt{numpy}), \code{Scipy} (\citealt{scipy}), \code{Astropy} (\citealt{astropy, astropy2_2018AJ....156..123A, astropy3_2022ApJ...935..167A}), \code{emcee} \citep{emcee_foremanmackey2013}, \code{Jupyter} Notebooks (\citealt{jupyter})}
%\cite{florian2020spatial}
%\cite{Khullar_2021}
%\cite{2000eaa..bookE1939B}

\appendix 
\section{Isochrones}
\label{appendix}
\begin{figure*}
    \centering
    \includegraphics[width=\textwidth]{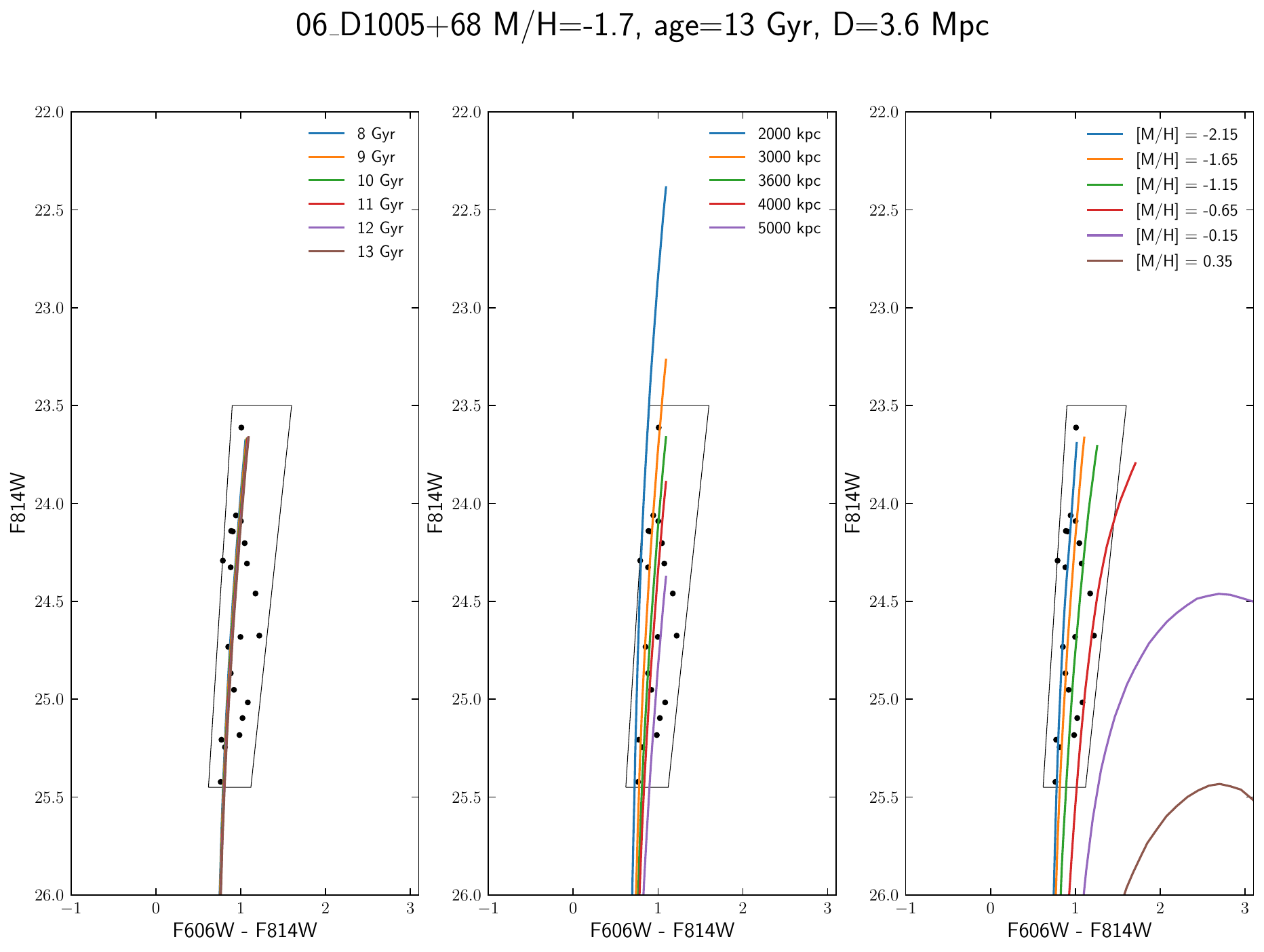}
    \caption{Various isochrones overlaid on the RGB stars within 2$a_h$ of D1005+68. The panels going from left to right show isochrones varying in age, distance, and metallicity, respectively, keeping the other two quantities constant with nominal values of $[M/H] = -1.5$, age$=13$ Gyr, and a distance of 3.6 Mpc.}
    \label{fig:dif_isochrones}
\end{figure*}

In Figure \ref{fig:dif_isochrones}, we show isochrones of varying age, distance, and metallicity overlaid on the RGB stars within 2$a_h$ of D1005+68. Varying the distance and metallicity of the isochrones (center and right panels, respectively) has a much larger effect on the isochrone fit than varying the age.

\section{Ellipticity vs. Separation}
\label{ellvsep}
\begin{figure*}[!h]
    \centering
    \includegraphics[width=\textwidth]{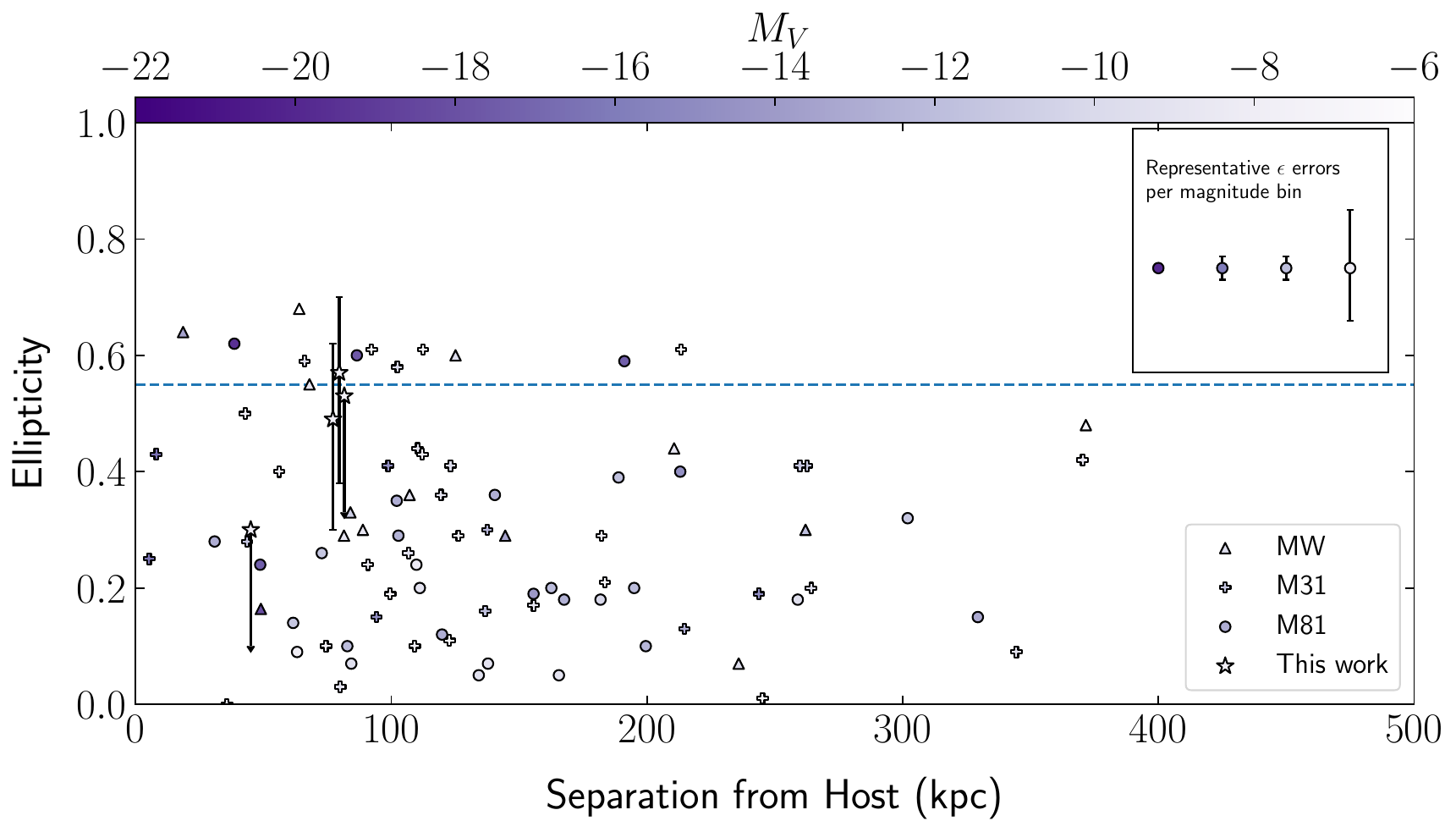}
    \caption{Same as Figure \ref{fig:ell_vs_tidal}, except the ellipticity of select Local Volume satellites against their separation from their host galaxy. M81 and M31 satellites are measured as their projected separations between M81 and M31, respectively. MW satellites are measured as their galactocentric 3D distances. The color map and errors are identical to the one used in Figure \ref{fig:ell_vs_tidal}. Data for MW and M31 satellites are from the Local Volume Database, data for M81 satellites are compiled from \cite{chiboucas2009_094469}.}
    \label{fig:ell_vs_sep}
\end{figure*}

Figure \ref{fig:ell_vs_sep} is similar to Figure \ref{fig:ell_vs_tidal}, except showing the separation from the host galaxy instead of the tidal ratio.

\section{Masks}
\label{masks}
\begin{figure*}[!h]
    \centering
    \includegraphics[width=0.49\textwidth]{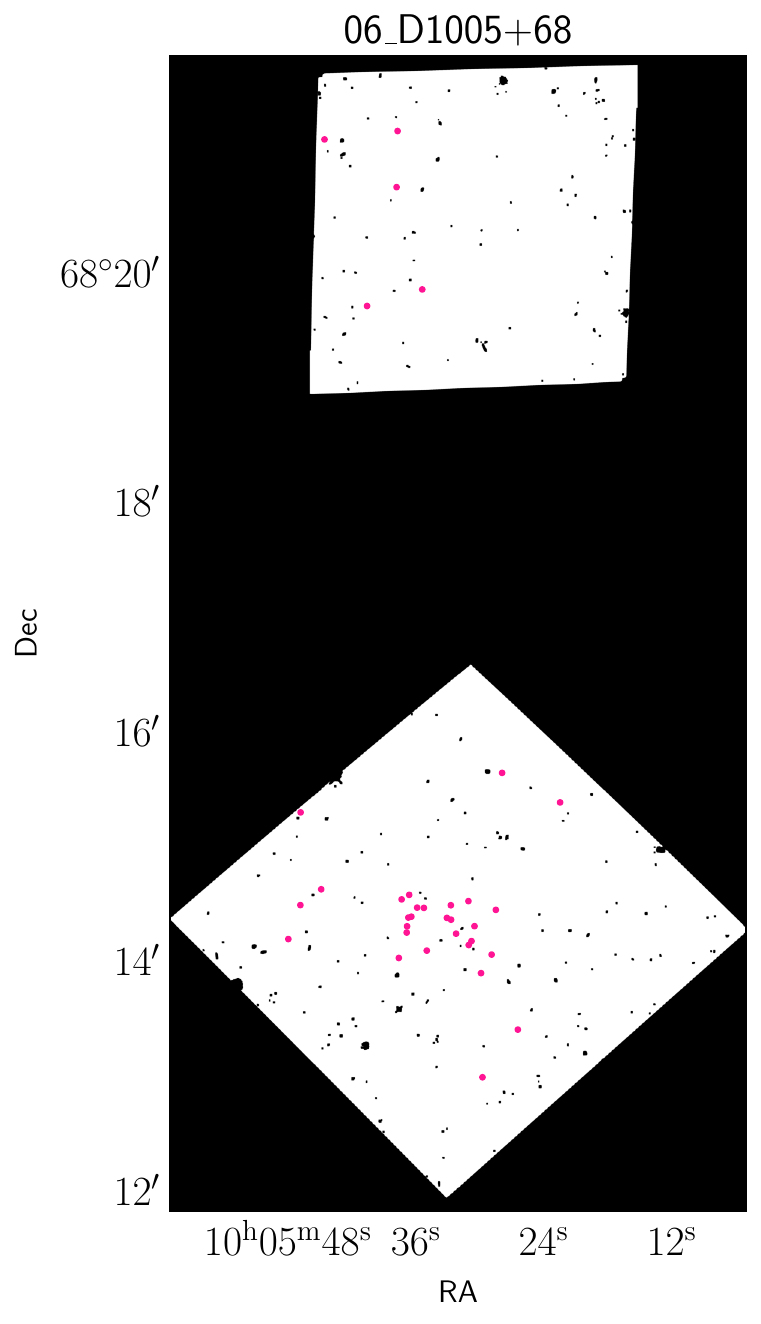}
    \includegraphics[width=0.49\textwidth]{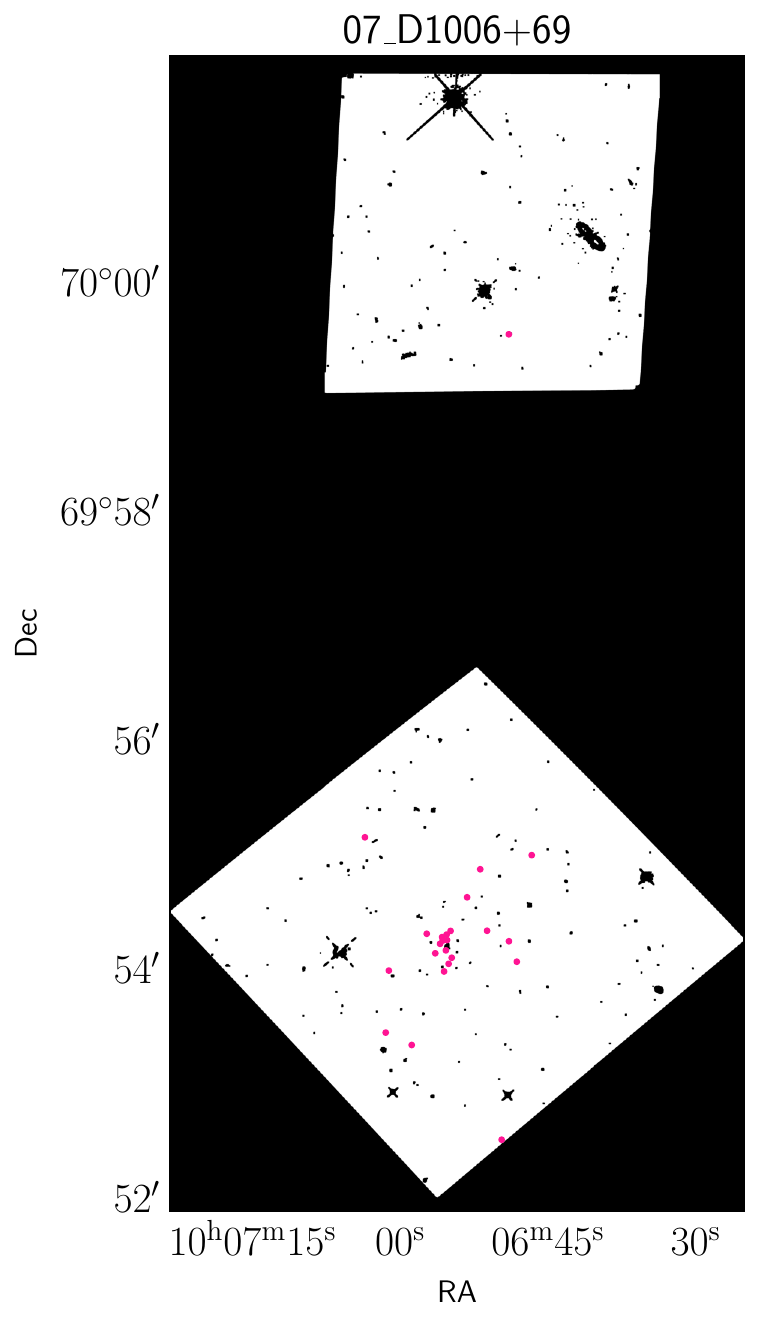}
    \includegraphics[width=0.49\textwidth]{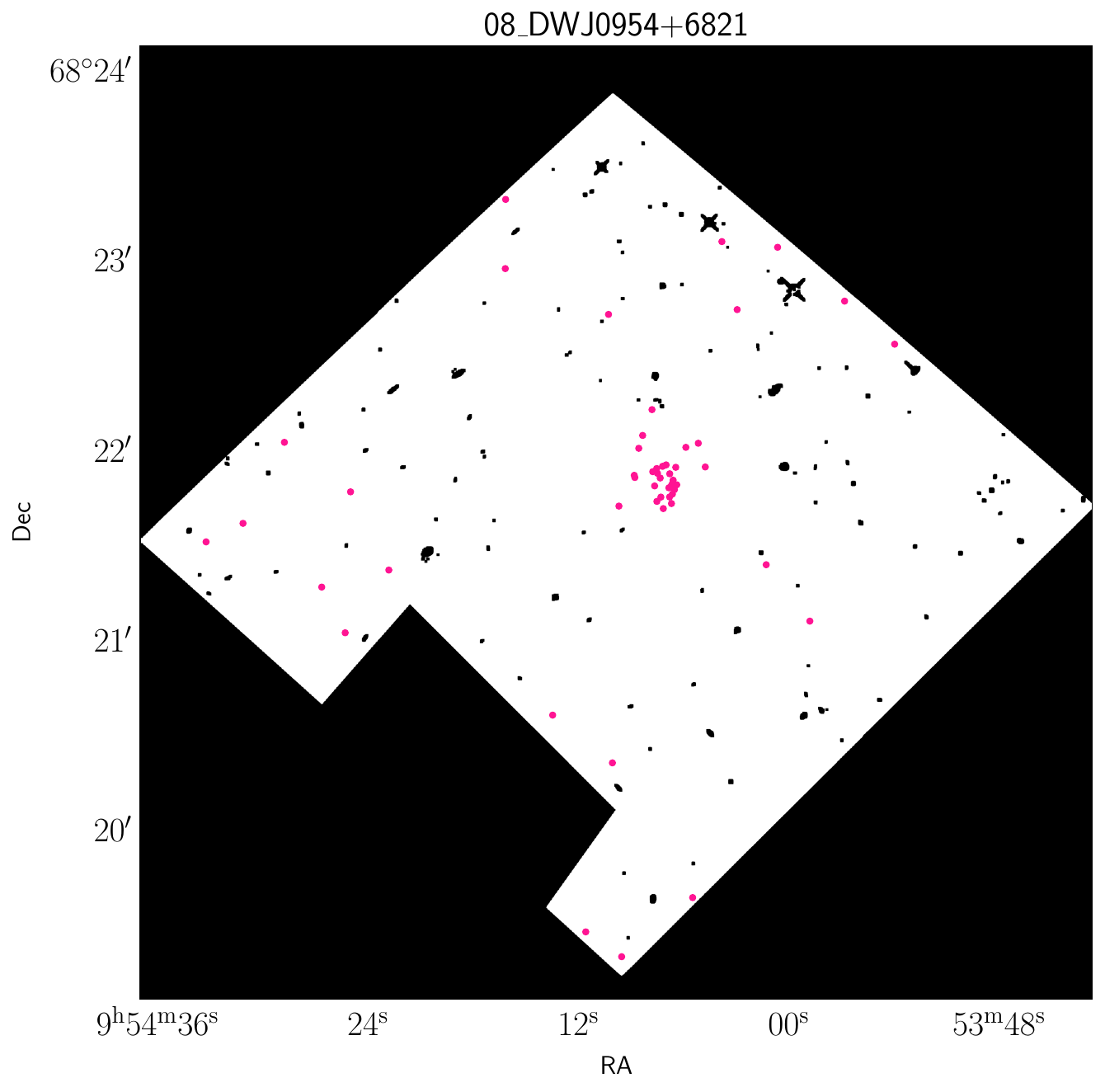}
    \includegraphics[width=0.49\textwidth]{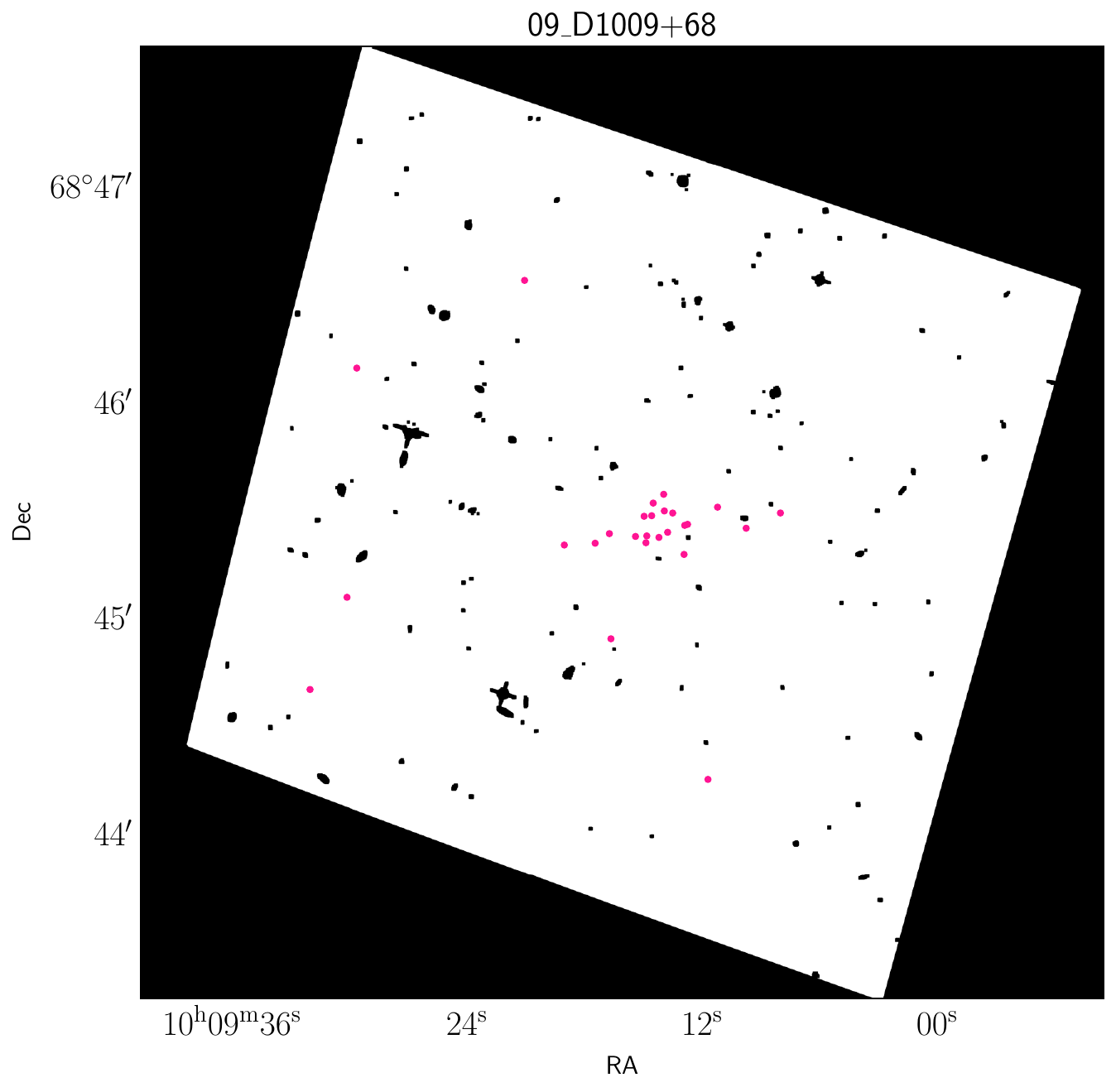}
    \caption{The mask images used to mask out bright stars and artifacts. RGB stars used for parameter fitting are overlaid as magenta points.}
    \label{fig:mask_images}
    
\end{figure*}

We display the masks and parallels (if applicable) used for each galaxy in Figure \ref{fig:mask_images}. Parallels were only used for \candsix and \candseven fitting to better constrain the background counts.

\newpage

\bibliography{sample631}{}
\bibliographystyle{aasjournal}



\end{document}